\documentclass[12pt]{article}

\usepackage[letterpaper,margin=2cm]{geometry}
\usepackage{graphicx}
\usepackage{amsmath}
\usepackage{amssymb}
\usepackage{comment}
\usepackage[hidelinks]{hyperref}
\usepackage{graphicx}
\usepackage{mathtools}
\usepackage{multirow}
\usepackage{titling}
\usepackage[nodisplayskipstretch]{setspace}
\usepackage{sectsty}
\usepackage[compact]{titlesec}
\usepackage{natbib}
\usepackage[labelfont=bf]{caption}
\usepackage{amsfonts}
\usepackage{newtxtext,newtxmath,microtype}
\usepackage[shortlabels]{enumitem}
\usepackage[table]{xcolor}  

\renewcommand{\thesection}{\arabic{section}}
\renewcommand{\thesubsection}{\arabic{section}.\arabic{subsection}}

\sectionfont{\normalfont\large\bfseries}
\subsectionfont{\normalfont\large\itshape}

\begin{document}

\pagenumbering{gobble}

\begin{center}
\Large
    Sampling distributions for complex design variance estimators in a Fay-Herriot model
\end{center}

\hspace{10pt}

\large
\noindent Alana McGovern$^{1*}$, Geir-Arne Fuglstad$^2$, and Jon Wakefield$^{1,3}$

\vspace{8pt}

\small  
\noindent$^1$ Department of Statistics, University of Washington\\
$^2$ Department of Mathematical Sciences, Norwegian University of Science and Technology\\
$^3$ Department of Biostatistics, University of Washington

\vspace{8pt}
\normalsize
\noindent $^*$Corresponding author, amcgov@uw.edu

\vspace{8pt}
\noindent Keywords: Bayesian small area estimation, variance smoothing, complex surveys

\clearpage

\doublespacing

\pagenumbering{arabic}

\begin{abstract}
\noindent Fay-Herriot (FH) models that jointly model the mean and variance typically use chi-squared sampling distributions for the design variance estimators, which require strong assumptions on the superpopulation model and are not fully compatible with complex survey designs, such as the stratified two-stage clustering design used by the Demographic and Health Surveys (DHS). DHS conducts surveys in low- and middle-income countries which result in low sample sizes for unplanned domains of interest. Hence, accounting for uncertainty in the estimated design variances is important.  We consider three sampling distributions for the design variance estimator: one which follows from simple random sampling; one which partially accounts for the DHS design; and one which fully accounts for the DHS design. In a simulation study, we compare FH models using these sampling distributions to a standard FH model. We find that both variance smoothing models which account for the complex design produce better credible intervals according to proper scoring rules. Interestingly, the sampling distribution which only partially accounts for survey design, and is easier to implement, performs equally as well as the sampling distribution which fully accounts for the survey design. We illustrate the proposed models by estimating height-for-age z-scores using the 2022 Kenya DHS.
\end{abstract}

\raggedright

\setlength{\parindent}{20pt}

\section{Introduction \label{sec:introduction}}

The Fay-Herriot area-level model is a common choice for small area estimation (SAE) because it accounts for the design by modeling survey-weighted estimates and enables borrowing of information across areas through covariates and random effects, which often increases precision and is particularly beneficial when areas have little or no data (\citealp{FHmodel}). In the sampling model for weighted means, the design variances of the weighted mean estimators are assumed to be known, but in practice, estimates of the variances are used. A common estimator of the design variance is obtained via Taylor linearization (\citealp{Wolter2007}, chap. 6; \citealp{korn_graubard}, chap. 2.4; \citealp{sarndal}, chap. 5.5; \citealp{lohr}, chap. 9.1). While the resulting estimators are design-consistent, they can be imprecise, particularly when sample size is small. A less-considered detriment is the tendency for this estimator to have downward bias, compared to the true design variance, for small sample sizes (\citealp{sarndal}, p. 176; \citealp{lohr}, chap. 9.1). As a result, taking design variance estimates as known in the Fay-Herriot model can result in interval estimates which are not only imprecise, but have systematic undercoverage for small sample sizes. 

This work is motivated by the challenges posed by estimating small area means using survey data in low- and middle-income countries (LMICs). The major surveys conducted in LMICs, particularly the Demographic and Health Surveys (DHS) (\citealp{dhs}) and Multiple Indicator Cluster Surveys (MICS) (\citealp{mics}), use stratified two-stage cluster designs where at the first stage a fixed number of clusters (enumeration areas) are sampled from each stratum using probability proportional to size (PPS) sampling, and at the second stage a fixed number of households are sampled from each cluster. These strata are generally chosen to be the first administrative (Admin-1) areas crossed with urban and rural classification, resulting in data which is powered to provide reliable survey-weighted estimates at the Admin-1 level.  See Appendix A of the 2022 Kenya DHS report \citep{kenya_report} for more details.  This design makes estimation for unplanned domains, for example, second administrative (Admin-2) areas, challenging because data is sparse. 

\begin{figure}[h]
    \centering
    \includegraphics[width=0.65\linewidth]{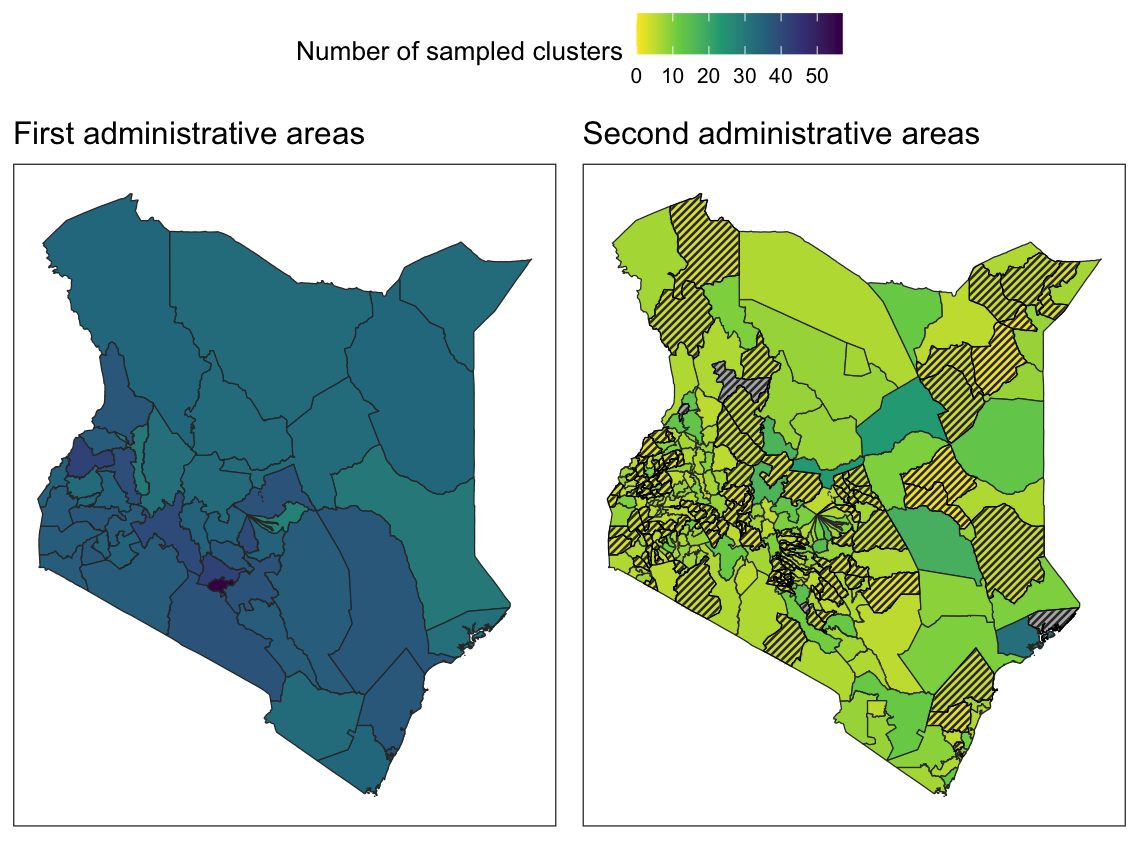}
    \caption{Number of sampled clusters in each Admin-1 and Admin-2 area in the 2022 Kenya DHS. Admin-2 areas with less than $5$ sampled clusters are marked by hatching and areas with no sampled clusters are filled in gray.}
    \label{fig:cluster_map}
\end{figure}

As an example, Figure \ref{fig:cluster_map} maps the number of clusters sampled from Admin-1 and Admin-2 areas in the 2022 Kenya DHS (\citealp{kenya_dhsdata}). The number of households in each cluster is not disclosed, but from the sampling frame we can deduce that each cluster contains an average of $94$ households. Design-based estimation for the $47$ Admin-1 areas is feasible, as they are planned domains with an average of $36$ clusters sampled per area. On average, $1.76\%$ of rural clusters and $3.8\%$ of urban clusters in each Admin-1 area are sampled. While the number of urban and rural clusters sampled from each Admin-1 area is fixed under the survey design, the number of clusters sampled from each Admin-2 area is random. In this survey, $6$ out of Kenya's $300$ Admin-2 areas have no sampled clusters and $127$ have less than five sampled clusters. As a result, there are several areas for which weighted mean estimates cannot be obtained and many for which estimates are imprecise or have unstable design variance estimates. observation, 

\begin{figure}[h!]
    \centering
    \includegraphics[width=0.85\linewidth]{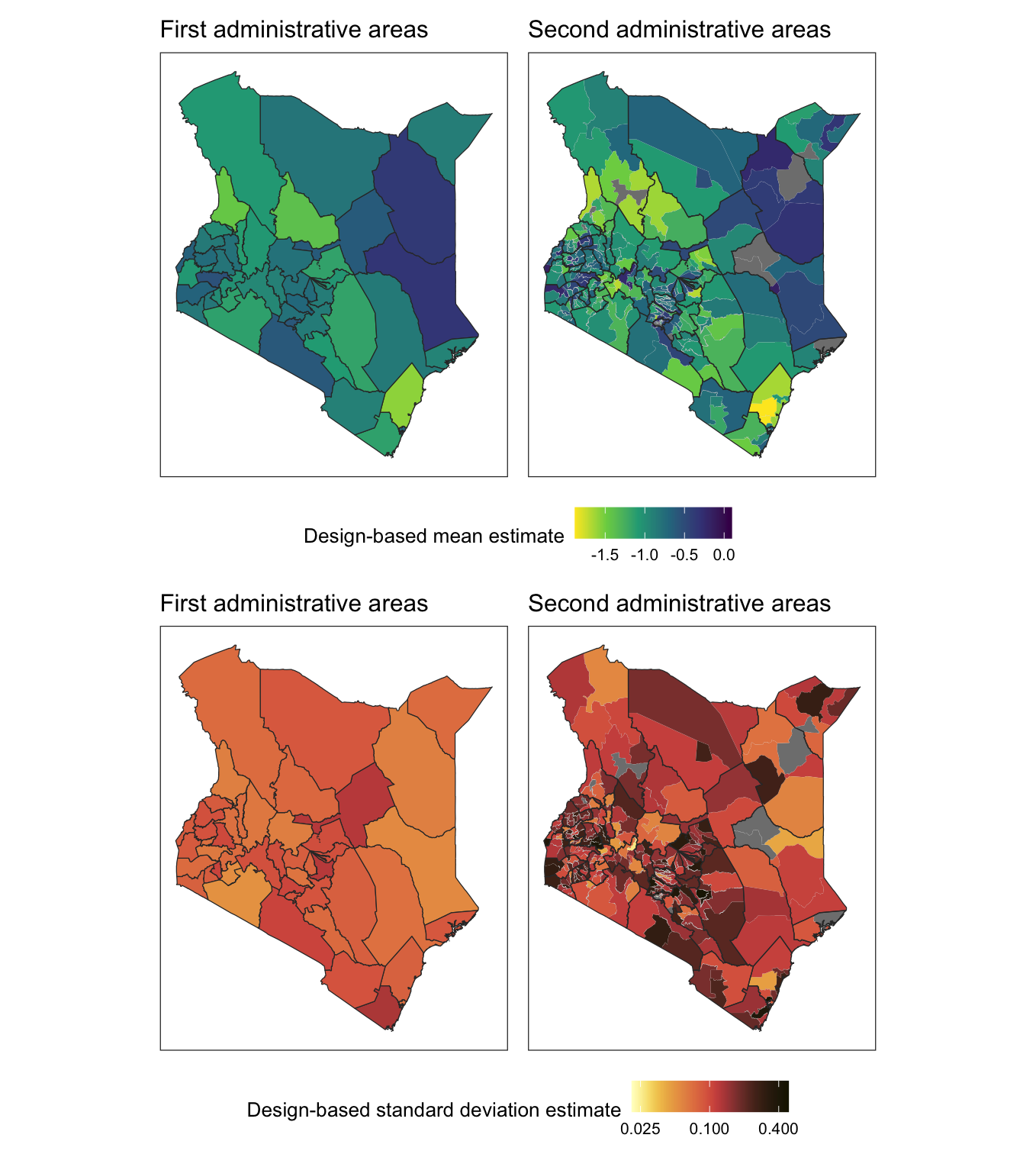}
    \caption{Design-based mean and standard deviation estimates for HAZ at the Admin-1 and Admin-2 levels using 2022 Kenya DHS.}
    \label{fig:kenya_direct}
\end{figure}

Consider the estimation of average height-for-age z-score (HAZ) for children under 5. The HAZ measures how many standard deviations a child's height is from the median height for children of the same sex and age, using median and standard deviations from the \cite{WHOgrowthstandards}. This measure is useful to evaluate child nutritional status relative to the international population. Reducing the rate of stunting, defined as a HAZ below $-2$, is one of the Sustainable Development Goals (SDGs) targeting malnutrition (\citealp{SDGs}). In Figure \ref{fig:kenya_direct} we map the design-based mean and standard deviation estimates for HAZ for children under 5, using the Kenya 2022 DHS. We observe notable between-area variation at the Admin-1 and Admin-2 levels. In this context it is particularly detrimental to use a Fay-Herriot model at the Admin-2 level which assumes the variances of the weighted estimates are known, as data sparsity yields variance estimates which are less precise and more likely to be biased. 

There is a vast literature concerning how variance estimates may be adjusted or have their uncertainty accounted for. When the outcome of interest is a proportion or total, one common strategy is to use a generalized variance function (GVF), which exploits the mean-variance relationship assumed under a binomial distribution (\citealp{valliant1987}; \citealp{Wolter2007}, chap. 7; \citealp{korn_graubard}, chap. 5.6; \citealp{lohr}, chap. 9.4). However, in the case of continuous outcomes, justifying any particular choice of GVF becomes difficult, as there is not a clear mean-variance relationship to exploit. Some work has been done by the US Bureau of Labor Statistics to use GVFs for time series analyses of continuous outcomes (\citealp{cho2002}; \citealp{mcillece2018}), but this requires frequent, data-rich surveys.

Another common strategy is to jointly model the sampling mean and sampling variance, so that the uncertainty of the design variance estimate is taken into account. This strategy requires specifying a sampling distribution for the design variance estimator. Previous work, including \cite{you2006}, \cite{maiti2014}, \cite{sugasawa2017}, \cite{erciulescu2019}, and \cite{parker2024}, assume that the survey is a stratified simple random sample with each area of interest, $i$, being a stratum, and assume individual outcomes within each stratum are identically and independently distributed (IID) normal. These assumptions result in the design variance estimator, $\hat V_i$, having the sampling distribution $V_i/\mathrm{df}_i\times \chi^2_{\mathrm{df}_i}$, where $V_i$ is the true design variance and the degrees of freedom ($\mathrm{df}_i$) are equal to the sample size minus one. 

\cite{gao2023spatial} modify this sampling distribution to account for a stratified cluster sampling design by using the appropriate nominal degrees of freedom, which is the number of sampled clusters minus the number of strata (\citealp{korn_graubard}, p. 34). As in \cite{gao2023spatial}, sampling distributions for complex design variance estimators have typically been employed in variance smoothing models without an examination of how fully they account for this complex design and the underlying population assumptions which must be made for these sampling distributions to hold. Sampling distributions for the design variance estimator can only be obtained in closed form under a restricted class of survey designs. In this work we will define a set of design properties and superpopulation assumptions under which this commonly used sampling distribution holds. We also introduce an alternative sampling distribution which fully accounts for the complex survey design under a defined superpopulation assumption and assess their performance in a Fay-Herriot model with complex survey data. We will test the robustness of these models to violations of the specified assumptions and compare them to the standard Fay-Herriot model which does not use variance smoothing and to a variance smoothing model whose sampling distribution does not account for the complex survey design.

We begin in Section \ref{sec:background} by introducing the two-stage cluster sampling design and defining the Fay-Herriot model using a superpopulation framework. In Section \ref{sec:derivations} we define superpopulation equivalents of two commonly used sampling distributions for the design variance estimators: one which assumes SRS and the other which partially accounts for a stratified two-stage cluster sampling design. We also a derive an alternative sampling distribution which fully accounts for the complex survey design. In Section \ref{sec:implementation} we fully specify our variance smoothing Fay-Herriot model using each of the proposed sampling distributions under a Bayesian framework. In Section \ref{sec:simulation_study} we conduct a simulation study in which we compare each proposed sampling distribution to the empirical distribution of the design variance estimator, as well as the model performance of a standard Fay-Herriot model (without variance smoothing) against Fay-Herriot models with variance smoothing using each of the proposed sampling distributions. Through this simulation study, we evaluate robustness of the models to violations of superpopulation assumptions and whether using a sampling distribution which more fully accounts for the complex survey design improves model performance. In Section \ref{sec:data_example} we apply variance smoothing models to HAZ data from the Kenya 2022 DHS and conclude with a discussion of our findings in Section \ref{sec:discussion}. 

\section{Fay-Herriot model for stratified two-stage cluster samples \label{sec:background}}

\subsection{Notation and setup \label{sec:setup}}

Consider a population composed of strata $\mathcal{H}=\{1,...,H\}$, where $H\geq 1$ is the number of strata. Stratum $h\in\mathcal{H}$, contains $M_h$ clusters, the set of which is denoted $U_h$. Cluster $c\in U=\cup_{h\in\mathcal{H}}U_h$ contains $N_c$ individuals. Let $y_{c\ell}$ denote the value of some continuous outcome for individual $\ell=1,...,N_c$ in cluster $c\in U$. Suppose we are interested in estimating the population means of this continuous outcome measure for a set of areas, $i=1,...,K$ wherein area $i$ spans the subset, $\mathcal{H}_i\subseteq\mathcal{H}$. Let $U_{hi}$ be the subset of clusters in $U_h$ which are contained in area $i$, with $M_{hi}=|U_{hi}|$. It follows that $U_{hi}=\emptyset$ for $h\in\mathcal{H}\setminus\mathcal{H}_i$. We take as the parameters of interest, the area-specific population means,  $$\theta_i=\frac{\sum_{h\in\mathcal{H}_i}\sum_{c\in U_{hi}}\sum_{\ell=1}^{N_c}y_{c\ell}}{\sum_{h\in\mathcal{H}_i}\sum_{c\in U_{hi}}N_c},\hspace{10mm} i=1,...,K.$$

Now consider a survey design, such as the DHS, in which at the first stage, $m_h\ll M_h$ clusters are sampled with PPS sampling based on the listed population size of the cluster, for $h\in\mathcal{H}$, the set of which is denoted $S_h\subset U_h$, and in the second stage, $n_c\ll N_c$ individuals are sampled with equal probability from each cluster $c\in S=\cup_{h\in\mathcal{H}}S_h$. Let the indicator variable, $\zeta_{cl}$, denote whether individual $\ell$ in cluster $c$ was sampled.  While the DHS samples households, rather than individuals, at the second stage, we abstract this detail because the inter- and intrahousehold correlation can be collapsed into within-cluster correlation \citep{Wolter2007,korn_graubard,lohr}. For the first stage, the sampling probability is determined by the listed size of cluster $c$, $L_c$, in a master sampling frame which is usually based on the most recent census. Thus the first stage sampling probability for cluster $c$ in stratum $h$, is $p_{1,c}=m_{h}L_c/\sum_{c'\in U_h}L_{c'}$. Clusters are defined to be of relatively similar size, which ensures $p_{1,c}<1$ for all $c\in U_h$. In practice, the true size of cluster $c$, $N_c$, will likely be different from that which is listed in the sampling frame, so in the second stage, the sampled clusters are re-enumerated and the second stage sampling probability is $p_{2,c}=n_c/N_c$. Then it follows the sampling weight for cluster $c$ in stratum $h$ is $$w_c=\frac{1}{p_{1,c} p_{2,c}}=\frac{N_c\sum_{c'\in U_h}L_{c'}}{L_cm_hn_c}.$$  As a result of this design, each individual in the same cluster has the same sampling weight, $w_c$. We assume throughout that $n_c\ll N_c$ and $m_h\ll M_h$, therefore the finite population corrections are negligible. 

Let $S_{hi}$ be the subset of clusters in $S_h$ which are contained in area $i$, with $m_{hi}=|S_{hi}|$, and let $m_{\cdot i}=\sum_{h\in\mathcal{H}_i}m_{hi}$ be the total number of sampled clusters in area $i$. Note that if area $i$ is a planned domain, $S_{hi}=S_h$ for all $h\in\mathcal{H}_i$, whereas if area $i$ is an unplanned domain, $S_{hi}\varsubsetneq S_h$, for $h\in\mathcal{H}_i$. Then the H\'ajek estimator (\citealp{hajek}) for $\theta_i$ is
\begin{equation}
    \hat \theta_i=\frac{\sum_{h\in\mathcal{H}_i}\sum_{c\in S_{hi}}w^\ast _c\bar y_c}{\sum_{h\in\mathcal{H}_i}\sum_{c\in S_{hi}}w^\ast _c}
    \label{eq:mean_estimator}
\end{equation}

\noindent where $\bar y_c=\left(\sum_{\ell}^{N_c}\zeta_{c\ell}y_{c\ell}\right)/n_c$ and $w^\ast _c=w_cn_c$. 

Following Equation 2.3-10 of \cite{korn_graubard}, we apply the Taylor linearization method to expression (\ref{eq:mean_estimator}) and obtain a design-consistent estimator for the design variance, $V_i=\mathrm{Var}(\hat\theta_i\mid\theta_i)$, conditioning on cluster sampling weights and sample sizes,
\begin{equation}
    \hat V_i=\frac{1}{\left(\sum_{h\in\mathcal{H}_i}\sum_{c\in S_h}w^\ast _c\right)^2}\sum_{h\in\mathcal{H}_i}\frac{m_h}{m_h-1}\sum_{c\in S_h}\left[w^\ast _c(\bar y_c-\hat \theta_i)-\frac{1}{m_h}\sum_{c'\in S_h}w^\ast _{c'}(\bar y_{c'}-\hat\theta_i)\right]^2
    \label{eq:variance_longform}
\end{equation}

\noindent where $w^\ast_c=0$ for all $c\in S_h\setminus S_{hi}$ and $h\in\mathcal{H}_i$, as these clusters are outside the area of interest. When the area is an unplanned domain we sum over the clusters in the larger planned domain to account for the additional variability introduced by the fact that the number of sampled clusters in the unplanned domain is random. See Appendix \ref{app:var_decomp} of the online Supplemental Materials for more details. Expression (\ref{eq:variance_longform}) is equivalent to the expression used in the \texttt{SURVEYMEANS} procedure in SAS \citep[p.~9282--9283]{sas}, and yields estimates which are nominally the same as those obtained using the \texttt{survey} package (\citealp{survey}) in R. If only one cluster is sampled from area $i$, $\hat\theta_i=\bar y_c$, so it follows that $\hat V_i=0$. While adjustments have been proposed to allow for variance estimation in this limiting case (\citealp{variance_fix}), we do not explore them here and assume that $m_{\cdot i}>1$. 

The standard Fay-Herriot model is
\begin{equation}
    \hat\theta_i\mid \theta_i \sim\mathcal{N}(\theta_i,\hat V_i),\hspace{10mm}\theta_i=\boldsymbol{X}_i^\mathrm{T}\boldsymbol{\beta}+b_i
    \label{eq:latent_mean}
\end{equation}

\noindent where $\boldsymbol{X}_i$ is an area-specific auxiliary variable vector, $\boldsymbol{\beta}$ is a vector of fixed effect regression parameters, and $\boldsymbol{b}=(b_1,...,b_K)^\mathrm{T}$ is a vector of the residual between-area variation, which are modeled as normal area-level random effects. In the original formulation of the Fay-Herriot model, the random effect is IID normal, but one may also include a spatially structured component in the random effect (\citealp{rao2015}, chap. 4.4.4; \citealp{chung2020}; \citealp{gao2023spatial}; \citealp{twocultures}).

\subsection{Variance smoothing Fay Herriot model \label{sec:var_smooth_setup}}

Now we will extend this framework to jointly model the design variances along with the means. It is common in the variance smoothing literature (\citealp{you2006}; \citealp{maiti2014}; \citealp{sugasawa2017}; \citealp{erciulescu2019}; \citealp{parker2024}) to use the sampling distribution, 
\begin{equation}
\hat V_i\mid V_i\sim \frac{V_i}{\mathrm{df}_i}\chi^2_{\mathrm{df}_i}
\label{eq:maiti_sampling}
\end{equation}

\noindent where df$_i=n_i^+-1$ and $n_i^+=\left(\sum_{h\in\mathcal{H}_i}\sum_{c\in S_{hi}}n_c\right)$. As \cite{maiti2014} note, this sampling distribution implies the data is IID normal and obtained via simple random sampling (SRS). \cite{you2006} give design variance, $V_i$, a diffuse inverse-gamma prior, which \cite{maiti2014} and \cite{sugasawa2017} note often leads to unstable estimation. The former modifies You and Chapman's model by taking an empirical Bayes approach, and the latter suggests using more informative hyperparameters and including covariates. \cite{erciulescu2019} chooses a normal distribution for $\log(V_i)$ conditional on covariates. A reasonable modification of (\ref{eq:maiti_sampling}) for the complex survey case, to account for stratification and clustering, is to simply plug in the appropriate nominal degrees of freedom for a complex survey, $\mathrm{df}_i=\sum_{h\in\mathcal{H}_i} \left(m_{hi}-1\right)$, as is done in \cite{gao2023spatial}.

In contrast to the above work which takes a design-based approach by directly modeling the design variance, $V_i$, we will use a superpopulation approach and model the superpopulation variance, $\sigma_i^2$, along the with the superpopulation mean, $\theta_i$. In particular, we assume 
\begin{equation}
    y_{c\ell}\mid\theta_{i},\gamma_h,\sigma_i^2\sim\mathcal{N}(\theta_{hi},\sigma_i^2),\hspace{10mm}\mathrm{for\hspace{1mm} all\hspace{1mm}}c\in U_{hi},
    \label{eq:pop_assumption}
\end{equation}

\noindent where $\theta_{hi}=\theta_i+\gamma_h$ is the superpopulation mean for strata $h$ and area $i$, $\left(\gamma_h\right)_{h\in\mathcal{H}}$ are stratum mean effects, and $\sigma_i^2$ is the within-stratum superpopulation variance for area $i$. These strata-specific means are well-defined under the Fay-Herriot model if $\gamma_h=0$ for one $h\in\mathcal{H}_i$ and we use the proportion of the population that belongs to each strata as an auxiliary variable in the latent mean model, as we will specify in Section \ref{sec:implementation}. 

We choose a latent model for $\sigma_i^2$ which mirrors that of the latent mean model, namely,\[
\mbox{log}(\sigma_i^2)=\boldsymbol{Z}^\mathrm{T}_i\boldsymbol{\eta} + g_i
\] where $\boldsymbol{Z}_i$ is an area-specific auxiliary variable vector, $\boldsymbol{\eta}$ is a vector of fixed effect regression parameters, and $\boldsymbol{g}=(g_1,...,g_K)$ is a vector of residuals, modeled as normal area-level random effects. This formulation allows for leveraging of any existing spatial structure, via the random effect and covariates, and an intuitive interpretation of the random effects and covariates.

\section{Sampling distributions for the complex design variance estimator \label{sec:derivations}}

Now we derive three candidate sampling distributions for the design variance estimator under superpopulation assumption (\ref{eq:pop_assumption}) and note which elements of the complex survey design described in Subsection \ref{sec:setup} are accounted for under each distribution.

\subsection{Simple random sampling-$\chi^2$ distribution}

First we present the simple random sampling-$\chi^2$ (SRS) distribution, which does not account for the complex survey design. Under SRS, it follows from superpopulation assumption (\ref{eq:pop_assumption}) that $V_i=\sigma_i^2/n_i^+$ where $n_i^+=\left(\sum_{h\in\mathcal{H}_i}\sum_{c\in S_{hi}}n_c\right)$. Moreover, an estimator for $V_i$ under SRS is $$\hat V_i=\frac{1}{n_i^+-1}\sum_{h\in\mathcal{H}_i}\sum_{c\in S_{hi}}\sum_{\ell}^{N_c}\left[\zeta_{c\ell}y_{c\ell}-\frac{1}{n_i^+}\sum_{h\in\mathcal{H}_i}\sum_{c\in S_{hi}}\sum_{\ell'}^{N_c}\zeta_{c\ell'}y_{c\ell'}\right]^2.$$  Then it follows from superpopulation assumption (\ref{eq:pop_assumption}) that

\begin{equation}
    \hat V_i\mid \sigma^2_i\sim \frac{\sigma^2_i/n_i^+}{n_i^+-1}\chi^2_{n_i^+-1}.
    \label{eq:srs_sampling}
\end{equation} 
Note that this is the superpopulation equivalent of the commonly used sampling distribution in (\ref{eq:maiti_sampling}).

\subsection{Complex sampling single-$\chi^2$ distribution \label{section:simple}}

Now we present a superpopulation equivalent of the sampling distribution used in \citet{gao2023spatial}. To derive this distribution, we assume the following:

\begin{enumerate}[(i)]
    \item Area $i$ is a planned domain (i.e., $S_{hi}=S_h$ and $m_{hi}=m_h$ for all $h\in\mathcal{H}_i$).
    \item Each stratum in area $i$ has the same number of sampled clusters (i.e., $m_{hi}=m_{\cdot i}/|\mathcal{H}_i|$).
    \item Each sampled cluster in area $i$ has the same number of sampled individuals (i.e., $n_c=n_0$).
    \item Each individual in area $i$ has the same sampling weight (i.e., $w_c=w_0$).
\end{enumerate}
Under these assumptions the variance estimator (\ref{eq:variance_longform}) simplifies to
\begin{equation}
\hat V_i =\frac{1}{m_{\cdot i}(m_{\cdot i}-|\mathcal{H}_i|)}\sum_{h\in\mathcal{H}_i}\sum_{c\in S_{hi}}\left[\bar y_c-\frac{1}{m_{hi}}\sum_{c'\in S_{hi}}\bar y_{c'}\right]^2.
     \label{eq:variance_simple}
\end{equation}

\noindent See Appendix \ref{app:simple_form} of the online Supplemental Materials for more details. Under superpopulation assumption (\ref{eq:pop_assumption}), it follows that $\bar y_c\sim\mathcal{N}(\theta_{hi},\sigma_i^2/n_0)$, for all $c\in S_{hi}$, $V_i=\sigma_i^2/n_i^+$, and
\begin{equation}
    \hat V_i\mid \sigma^2_i\sim\frac{\sigma_i^2/n_i^+}{m_{\cdot i}-|\mathcal{H}_i|}\sum_{h\in\mathcal{H}_i}\chi^2_{m_{hi}-1} = \frac{\sigma_i^2/n_i^+}{m_{\cdot i}-|\mathcal{H}_i|}\chi^2_{m_{\cdot i}-|\mathcal{H}_i|}.
    \label{eq:simple_approx}
\end{equation}

\noindent We refer to this distribution as the complex sampling single-$\chi^2$ (CSS) distribution because it partially accounts for the complex survey design described in Subsection \ref{sec:setup} using a single $\chi^2$ variable. While this distribution accounts for the complex survey design by considering the number of sampled strata and clusters, it does not account for differences in sample size and sampling weights across strata and clusters, or whether the area is a planned domain. In Section \ref{sec:simulation_study} we will use simulations to study whether not accounting for these elements of the complex survey design impacts model estimates.

As a consequence of assuming cluster sample sizes are equal, we can relax the assumption of independence within clusters and allow for all clusters in area $i$ to have the same within-cluster correlation. To illustrate this, suppose that individual outcomes in cluster $c\in U_{hi}$ are normally distributed with mean, $\theta_{hi}$, variance $\sigma_i^2$, and a correlation of $\rho_0$ between all individuals in cluster $c$. Then it follows that $$\mathrm{Var}(\bar y_c)=\sigma_i^2\left(1+\rho_0\left(n_0-1\right)\right)/n_0,$$ so the within-cluster correlation is absorbed into the estimation of $\sigma_i^2$ because $n_0$ is a constant across clusters.

\subsection{Complex sampling mixture-$\chi^2$ distribution \label{satt_section}}

In this section we derive a sampling distribution for (\ref{eq:variance_longform}) which, in contrast to the CSS distribution, fully accounts for the survey design described in Subsection \ref{sec:setup}. Under superpopulation assumption (\ref{eq:pop_assumption}) and SRS within each cluster it follows that 
\begin{equation}
    \boldsymbol{\bar y_i}\mid \theta_i,\boldsymbol{\gamma_i},\sigma_i^2 \sim \mathcal{N}\left(\theta_i\boldsymbol{1}+\boldsymbol{\gamma_i},\sigma_i^2 {\bf D}_{\boldsymbol{i}}\right),
    \label{eq:cluster_distribution}
\end{equation}
where $\boldsymbol{\bar y_i}$ is composed of subvectors, $(\boldsymbol{\bar y}_{\boldsymbol{i}h})_{h\in\mathcal{H}_i}$, of length $m_h$ with elements $(\bar y_c)_{c\in S_h}$, $\boldsymbol{\gamma_i}$ is a vector composed of subvectors, $\left(\gamma_h\boldsymbol{1}_{m_h}\right)_{h\in\mathcal{H}_i}$, and ${\bf D}_{\boldsymbol{i}}$ is a a diagonal matrix with elements $(1/n_c)_{c\in \cup_{h\in\mathcal{H}_i}S_h}$.

Let ${\bf B}_{\boldsymbol{i}}$ be a block diagonal matrix with blocks $\left({\bf B}_h\right)_{h\in\mathcal{H}_i}$ such that ${\bf B}_h=\frac{m_h}{m_h-1}\left({\bf I}_{m_h}-\frac{1}{m_h}{\bf 1}_{m_h}{\bf 1}^\mathrm{T}_{m_h}\right)$ where ${\bf 1}_{m_h}$ is an $m_h-$length column vector of $1$s. Let $\boldsymbol{w_i}^\ast$ be a vector composed of subvectors, $(\boldsymbol{w}^\ast_h)_{h\in\mathcal{H}_i}$, with elements $\left(w^\ast _c\right)_{c\in S_h}$ and define ${\bf T}_{\boldsymbol{i}}={\bf W}_{\boldsymbol{i}}\left({\bf I}_{m_{\cdot i}}-\frac{{\bf 1}\boldsymbol{w_i}^{\ast \mathrm{T}}}{{\bf 1}^\mathrm{T}\boldsymbol{w_i}^\ast }\right)$, where ${\bf W}_{\boldsymbol{i}}=\mbox{diag}\left(\boldsymbol{w_i}^\ast \right)$. Note that $(\boldsymbol{\bar y_i},\boldsymbol{\gamma_i},{\bf D}_{\boldsymbol{i}},{\bf B}_{\boldsymbol{i}},\boldsymbol{w_i}^\ast)$ must maintain consistent ordering. Then we can express the weighted mean estimator (\ref{eq:mean_estimator}) and its corresponding variance estimator (\ref{eq:variance_longform}) as,
\begin{equation}
\hat\theta_i=\frac{\boldsymbol{w_i}^{\ast \mathrm{T}}\boldsymbol{\bar y_i}}{{\bf 1}^\mathrm{T}\boldsymbol{w_i}^\ast }, \hspace{30mm}  \hat V_i=\frac{1}{\left({\bf 1}^\mathrm{T}\boldsymbol{w_i}^\ast \right)^2}\boldsymbol{\bar y_i}^\mathrm{T}{\bf M}_{\boldsymbol{i}}\boldsymbol{\bar y_i}
    \label{eq:matrixform}
\end{equation}

\noindent where ${\bf M}_{\boldsymbol{i}}={\bf T}_{\boldsymbol{i}}^\mathrm{T}{\bf B}_{\boldsymbol{i}}{\bf T}_{\boldsymbol{i}}$. See Appendix \ref{matrix_form_appendix} of the online Supplemental Materials for more details regarding the equality of the form of the variance estimator expressed in (\ref{eq:variance_longform}) and the matrix form expressed in (\ref{eq:matrixform}). 

Let $r_i\leq m_{\cdot i}$ denote the rank of ${\bf D}_{\boldsymbol{i}}^{1/2}{\bf M}_{\boldsymbol{i}}{\bf D}_{\boldsymbol{i}}^{1/2}$, $\boldsymbol{q_i}$ denote the $r_i$-dimensional vector of non-zero eigenvalues of ${\bf D}_{\boldsymbol{i}}^{1/2}{\bf M}_{\boldsymbol{i}}{\bf D}_{\boldsymbol{i}}^{1/2}$, and ${\boldsymbol{v}_{\boldsymbol{i}j}}$ denote the $m_{\cdot i}$-length eigenvector corresponding to eigenvalue $q_{ij}$, scaled such that $\boldsymbol{v}_{\boldsymbol{i}j}^\mathrm{T}{\bf D}_{\boldsymbol{i}}\boldsymbol{v}_{\boldsymbol{i}j}=1$. Then under (\ref{eq:cluster_distribution}), we obtain a sampling distribution for $\hat V_i$, as a sum of scaled chi-squares, namely,
\begin{equation}
    \hat V_i\mid \boldsymbol{\gamma_i},\sigma_i^2\sim\frac{\sigma_i^2}{\left({\bf 1}^\mathrm{T}\boldsymbol{w_i}^\ast \right)^2}\sum_{j=1}^{r_i}q_{ij}\chi^2_1(\delta_{ij})
    \label{eq:exact_dist1}
\end{equation}
where $\chi^2_1(\delta_{ij})$ denotes a chi-square distribution with $1$ degree of freedom and noncentrality parameter, $\delta_{ij}=\left({\boldsymbol{v}_{\boldsymbol{i}j}}^\mathrm{T}\boldsymbol{\gamma_i}\right)^2/\sigma_i^2$. It follows from (\ref{eq:matrixform}) that under superpopulation assumption (\ref{eq:pop_assumption}) and the survey design described in Subsection \ref{sec:setup},  $V_i=\sigma_i^2/n_i^\dagger$ where 
\begin{equation}
    n_i^\dagger = \frac{\left({\bf 1}^\mathrm{T}\boldsymbol{w_i}^\ast \right)^2}{\boldsymbol{w_i}^{\ast \mathrm{T}}{\bf D}_{\boldsymbol{i}}\boldsymbol{w_i}^{\ast }}.
    \label{eq:e_star}
\end{equation}
 The sampling distribution defined in (\ref{eq:exact_dist1}), which we will refer to as the complex sampling mixture-$\chi^2$ (CSM) distribution, can also be expressed as
\begin{equation}
    \hat V_i\mid \boldsymbol{\gamma_i},\sigma^2_i\sim\frac{\sigma_i^2/n_i^\dagger}{\boldsymbol{w_i}^{\ast \mathrm{T}}{\bf D}_{\boldsymbol{i}}\boldsymbol{w_i}^{\ast }}\sum_{j=1}^{r_i}q_{ij}\chi^2_1(\delta_{ij}).
    \label{eq:exact_dist}
\end{equation}

For ease of implementation in the Fay-Herriot model, it is preferable to approximate the sum of non-central chi-square random variables of (\ref{eq:exact_dist}) with a single random variable. The Satterthwhaite approximation is a single scaled chi-square random variable which matches the first two moments of a sum of scaled chi-squares (\citealp{satterthwaite}). The mean and variance of $\sum_j^{r_i}q_{ij}\chi^2_1(\delta_{ij})$ are $Q_1=\sum_j^{r_i} q_{ij}(1+\delta_{ij})$ and $Q_2=2\sum_j^{r_i} q_{ij}^2(1+2\delta_{ij})$, respectively, so by applying this moment-matching approach, we obtain an approximation of the CSM distribution,
\begin{equation}
\hat V_i\mid \boldsymbol{\gamma_i},\sigma^2_i \sim \frac{\sigma^2_i/n_i^\dagger}{\boldsymbol{w_i}^{\ast \mathrm{T}}{\bf D}_{\boldsymbol{i}}\boldsymbol{w_i}^{\ast }}\frac{Q_2}{2Q_1}\chi^2_{2Q_1^2/Q_2}.
\label{eq:satt_approx}
\end{equation}

\noindent In \ref{appendix:satt_approx} of the online Supplemental Materials we illustrate via several examples that this approximation has high accuracy. Note that $\hat V_i\mid \boldsymbol{\gamma_i},\sigma_i^2$ is a biased estimator for the theoretical design variance, $\sigma^2_i/n_i^\dagger$, under this sampling distribution. We show in \ref{appendix:bias_derivation} of the online Supplemental Materials that, under mild design assumptions, this bias is downward (i.e., $\mathbb{E}[\hat V_i\mid \sigma_i^2,\boldsymbol{\gamma_i}]-\sigma^2_i/n_i^\dagger<0$), which reflects the fact that while $\hat V_i$ is a consistent estimator for the design variance, it is generally a downwardly-biased estimator for finite samples (\citealp{sarndal}, p.176; \citealp{lohr}, p. 369).

\section{Variance smoothing Fay-Herriot model specification\label{sec:implementation}}

Let $\boldsymbol{\theta}=(\theta_1,...,\theta_K)$ and $\boldsymbol{\sigma}^2=(\sigma_1^2,...,\sigma^2_K)$ denote the superpopulation means and within-stratum superpopulation variances for areas $i=1,...,K$. For each area $i$, we obtain a H\'ajek estimate for $\theta_i$ from expression (\ref{eq:mean_estimator}), denoted $\hat\theta_i$, and a design variance estimate from expression (\ref{eq:variance_longform}), denoted $\hat V_i$. We define a Fay-Herriot model with variance smoothing
\begin{equation}
\begin{aligned}
\hat{\theta}_i \mid \theta_i, \sigma_i^2 
&\sim \mathcal{N}(\theta_i, V_i(\sigma_i^2)), 
&\quad
\hat{V}_i \mid \gamma, \sigma_i^2 
&\sim V_i(\sigma_i^2)\, \frac{c_i}{d_i(\gamma, \sigma_i^2)}\, \chi^2_{d_i(\gamma,\sigma_i^2)}, \\
\theta_i 
&= \boldsymbol{X}_i^\mathrm{T} \boldsymbol{\beta} + p_i \gamma + b_i, 
&\quad
\log(\sigma_i^2) 
&= \boldsymbol{Z}_i^\mathrm{T} \boldsymbol{\eta} + g_i,
\end{aligned}
\label{eq:model_spec}
\end{equation}

\noindent where $\boldsymbol{X}_i$ and $\boldsymbol{Z}_i$ are area-specific auxiliary variable vectors, $(\boldsymbol{\beta},\gamma)$ and $\boldsymbol{\eta}$ are vectors of fixed effect regression parameters, $p_i$ is the urban population proportion, and $\boldsymbol{b}$ and $\boldsymbol{g}$ are vectors of the residual between-area variation modeled as normal area-level random effects, described in more detail below.  The functions $V_i(\sigma_i^2)$ and $d_i(\gamma,\sigma^2_i)$ and constant $c_i$ are defined to attain the SRS, CSS or CSM distribution. For the SRS distribution,
\begin{equation}
    V_i(\sigma_i^2)=\frac{\sigma_i^2}{n_i^+},\quad c_i=1,\quad \mathrm{and} \quad d_i(\gamma,\sigma^2_i)=n_i^+-1,
    \label{eq:srs_theoretical_var}
\end{equation} for the CSS distribution (\ref{eq:simple_approx}),
\begin{equation}
    V_i(\sigma_i^2)=\frac{\sigma_i^2}{n_i^+},\quad c_i=1,\quad \mathrm{and} \quad d_i(\gamma,\sigma^2_i)=m_{\cdot i}-|\mathcal{H}_i|,
    \label{eq:simple_theoretical_var}
\end{equation}
and for the CSM distribution (\ref{eq:satt_approx}), 
\begin{equation}
    V_i(\sigma_i^2)=\frac{\sigma_i^2}{n_i^\dagger},\quad c_i=\frac{1}{\boldsymbol{w_i}^{\ast \mathrm{T}}{\bf D}_{\boldsymbol{i}}\boldsymbol{w_i}^{\ast}}, \quad \mathrm{and}
    \quad d_i(\gamma,\sigma^2_i)=\frac{2Q_1(\gamma,\sigma_i^2)^2}{Q_2(\gamma,\sigma_i^2)},
    \label{eq:sasw_theoretical_var}
\end{equation} where $n_i^\dagger$ is as defined in (\ref{eq:e_star}) and $Q_1(\gamma,\sigma_i^2)$ and $Q_2(\gamma,\sigma_i^2)$ are as defined in (\ref{eq:satt_approx}). 

The area-level random effects, $\boldsymbol{b}$ and $\boldsymbol{g}$, can either be IID normal with precision hyperparameters $\tau_b$ and $\tau_g$, respectively, or the BYM2 spatial model with hyperparameters $(\tau_b,\phi_b)$ and $(\tau_g,\phi_g)$, respectively.  The BYM2 model, introduced by \cite{BYM2}, is a re-parameterized version of the Besag-York-Molli\'e (BYM) model (\citealp{BYM}), which includes both unstructured IID normal random effects and structured, intrinsically conditional autoregressive (ICAR), spatial random effects (\citealp{BYM}). The BYM2 model has two parameters: $\tau$, indicating the overall precision of the random effect, and $\phi$, indicating the proportion of variation that is explained by the structured component. The structured component  has a sum-to-zero constraint to ensure identifiability and is scaled such that the marginal standard deviation of the random effect is approximately $1/\sqrt{\tau}$ (\citealp{BYM2}). 

For $(\boldsymbol{\eta},\boldsymbol{\beta},\gamma)$ we use normal priors with fixed mean and variance. For the precision parameters of the BYM2 random effects, $\boldsymbol{b}$ and $\boldsymbol{g}$, we use penalized complexity (PC) priors \citep{PCPriors} which place a constant-rate penalty on the Kullback-Leibler divergence between a flexible model and its simpler base model where, in this context, the base model is one with no random effect $(\tau=\infty)$. On the precision scale, this corresponds to a Type-2 Gumbel density with parameter $\lambda$. The rate $\lambda$ is calibrated by specifying a tail probability, $P(1/\sqrt{\tau}>u) = \alpha$.

\section{Simulation study \label{sec:simulation_study}}

In this study we assess the importance of accounting for the uncertainty of the design variance estimator by comparing the performance of Fay-Herriot models with variance smoothing using the SRS, CSS, and CSM distributions to the performance of the standard Fay-Herriot model with no variance smoothing. In addition to testing robustness against violations of superpopulation assumption (\ref{eq:pop_assumption}), we evaluate how model performance is affected by using a sampling distribution which fully (CSM) or partially (CSS) accounts for the complex survey design, as opposed to using the SRS distribution which does not properly account for survey design.  We also compare the performance of these variance smoothing models when informative covariates for the variance are unavailable and the variance is not spatially structured. The focus of our study is estimating means and their uncertainty at the Admin-2 level, which are unplanned domains. 

\subsection{Simulation design \label{sec:sim_design}}

To mimic our motivating context we use the geography of Kenya, which has 47 Admin-1 areas and $K=$ 300 Admin-2 areas and choose simulation settings which mimic the sampling frame and design of the 2022 Kenya DHS (\citealp{kenya_report}). For the sampling frame, we use the total number of clusters in each stratum $h$, $M_h$, provided in the 2022 Kenya DHS report. The number of clusters in the sampling frame of each Admin-2 area is not reported, so we group clusters within each stratum into their nested Admin-2 areas so that the number of clusters is proportional to its total population, for which we use WorldPop estimates (\citealp{worldpop}). Each Admin-2 area has the under-5 population estimated by WorldPop and each individual is assigned to a cluster in its respective area with equal probability. 

The survey uses a stratified two-stage cluster design with $H=92$ strata: an urban and rural stratum for each Admin-1 area, except for two areas which are solely urban. Let $\mathcal{H}^\mathrm{U}\subset\mathcal{H}$ and $\mathcal{H}^\mathrm{R}\subset\mathcal{H}$ denote the subsets of urban and rural strata, respectively. For each stratum $h$, we draw a fixed number of clusters, $m_h$, with PPS sampling, which is summarised in Figure \ref{fig:cluster_map} (except for one simulation setting, noted below, in which we increase the number of sampled clusters). We assume the listed size of each cluster in the sampling frame is correct (i.e., $L_c=N_c$) and, therefore, we do not require re-enumeration at the second stage. See Appendix \ref{appendix:sim_additional} of the online Supplemental Materials for an additional simulation setting which considers re-enumeration of cluster sizes. 

While the DHS samples a fixed number of households from each cluster, we directly sample individuals from each cluster, with the distribution of sample sizes across clusters being chosen to closely emulate the sample sizes in the 2022 Kenya DHS. For each urban cluster, the number of sampled individuals, $n_c$, is drawn from a negative binomial distribution with size, $8$, and mean, $9$. For each rural cluster, the number of sampled individuals, $n_c$, is drawn from a negative binomial distribution with size, $4$, and mean, $11$. See \ref{sample_size_appendix} of the online Supplemental Materials for more details. The sampling weight for an individuals in cluster $c\in S_h$ is the inverse of their sampling probability so, according to this design, $w_c=(\sum_{c'\in U_h}N_{c'})/(m_{h}n_c)$, where $\sum_{c'\in U_h}N_{c'}$ is the population size in stratum $h$. 

We define the superpopulation mean and within-stratum superpopulation variance for area $i$ in stratum $h$ as 
\begin{equation}
    \theta_{hi}=\boldsymbol{X}^\mathrm{T}_i\boldsymbol{\beta} + \gamma_i\mathbb{I}\left(h\in\mathcal{H}^U\right)+b_i,\hspace{5mm} \mbox{log}(\sigma_{hi}^2)=\boldsymbol{Z}^\mathrm{T}_i\boldsymbol{\eta}+\kappa_i\mathbb{I}\left(h\in\mathcal{H}^U\right)+g_i,
    \label{eq:sim_mean_var}
\end{equation}

\noindent where $\boldsymbol{X}_i$ and $\boldsymbol{Z}_i$ are each vectors of length $4$ with the first element equal to $1$ and the remaining elements having values generated from a standard normal distribution. We choose $\boldsymbol{\beta}=(-1,-0.15,-0.1,0.25)^\mathrm{T}$, $\boldsymbol{\eta}=(0.5,-0.15,-0.1,0.25)^\mathrm{T}$, and $\boldsymbol{\gamma}$ and $\boldsymbol{\kappa}$ are area-specific urban effects which vary across settings. Note that the simulated covariates are different for the mean and variance latent models. The between-area variation is quantified by area-level random effects, $\boldsymbol{b}=(b_1,...,b_K)$ and $\boldsymbol{g}=(g_1,...,g_K)$ which are generated from independent BYM2 models with parameters $(1/\sqrt{\tau_b}\approx0.32,\phi_b=0.75)$ and $(1/\sqrt{\tau_g}\approx0.22,\phi_g=0.75)$. Given these choices of $(\boldsymbol{\beta},\boldsymbol{\eta},\tau_b,\tau_g,\phi_b,\phi_g)$, approximately 50\% of the between-area variability in the mean and within-stratum variance can be explained through the covariates, and 75\% of the remaining 50\% can be explained through the spatial structure. 

Five simulation settings, summarized in Table \ref{tab:settings}, are used to assess how the models perform under correct specification, increased sample size, and misspecification of the latent models and superpopulation distribution. In setting 1, we simulate data to be correctly specified under superpopulation assumption (\ref{eq:pop_assumption}) and the model defined in (\ref{eq:model_spec}). Specifically, we choose $\kappa_i=0$ and $\gamma_i=1$ for each area $i$, which restricts the model to have equal within-stratum superpopulation variance within each area $(\sigma_{hi}^2=\sigma_i^2)$, and equal difference between stratum means among all areas $(\gamma_i=\gamma)$. Then for each individual $\ell$ in cluster $c\in U_{hi}$, we generate $y_{c\ell}\sim\mathcal{N}(\theta_{hi},\sigma_{hi}^2)$. In setting 1a we generate outcomes in the same manner, but sample five times as many clusters as in the 2022 Kenya DHS.

\begin{table}[h]
    \centering
    \caption{Summary of simulation settings, where $\theta_{hi}$ and $\sigma^2_{hi}$ are as defined in (\ref{eq:sim_mean_var})}
    \begin{tabular}{l|l|c|c}
        Setting & Description & Urban effects & Superpopulation distribution \\
        \hline\hline
        1 & correctly specified & $\gamma_i=1;\kappa_i=0$ & $\mathcal{N}(\theta_{hi},\sigma_{hi}^2)$\\
        \hline
        1a & correctly specified & $\gamma_i=1;\kappa_i=0$ & $\mathcal{N}(\theta_{hi},\sigma_{hi}^2)$\\
        & with $5$ times more& & \\
        & sampled clusters & & \\
        \hline
        2 & varying urban & $\gamma_i\sim\mathcal{N}(1,1^2)$ & $\mathcal{N}(\theta_{hi},\sigma_{hi}^2)$\\
        & effects & $\kappa_i\sim\mathcal{N}(\log (1.5),0.25^2)$ & \\
        \hline
        3 & t-distributed data & $\gamma_i=1;\kappa_i=0$ & t-distributed with mean $\theta_{hi}$,\\
        & & & variance $\sigma_{hi}^2$, and df$=5$\\
        \hline
        4 & within-cluster & $\gamma_i=1;\kappa_i=0$ & normal with mean $\theta_{hi}$,\\
        & correlation &  & variance $\sigma_{hi}^2$, and within-cluster\\
        & &  & correlation, $\rho=0.25$\\
    \end{tabular}
    \label{tab:settings}
\end{table}

In setting 2, we test misspecification of the latent mean and variance models. We generate outcomes in the same manner as in setting 1, with the only difference being that we draw $\gamma_i\sim\mathcal{N}(1,1^2)$ and $\kappa_i\sim\mathcal{N}(\log(1.5),0.25^2)$, which implies that the within-stratum superpopulation variance in an area is $50\%$ higher, on average, in the urban stratum than in the rural stratum. In settings 3 and 4, we test violations of superpopulation assumption (\ref{eq:pop_assumption}). Both settings use the same latent models as in setting 1. In setting 3, we relax the normality assumption and generate independent outcomes from a t-distribution with 5 degrees of freedom, shifted and scaled so that the mean and variance is $\theta_{hi}$ and $\sigma^2_{hi}$, respectively. In setting 4, we relax the assumption of independence within clusters by generating outcomes from a normal distribution with mean $\theta_{hi}$, variance $\sigma^2_{hi}$, and a correlation of $\rho=0.25$ within each cluster. For each of the $5$ settings we generate a fixed population surface and sample $100$ datasets from this population surface using the stratified two-stage cluster sampling design described above. A summary of Admin-2 area sample sizes is provided in Table \ref{tab:sim_samplesizes}. 

\begin{table}[h]
    \centering
    \caption{Summary of Admin-2 level sample sizes across 100 simulations for each setting.}
    \begin{tabular}{l|c|c}
         & Settings 1-4 & Setting 1a \\
         & Mean (SD) & Mean (SD) \\
         \hline\hline
        Number of sampled clusters: &  & \\
         \hspace{0.5em} Urban & 2.2 (2.0) & 11.2 (8.2) \\
         \hspace{0.5em} Rural & 3.4 (2.7) & 17.0 (11.3) \\
         \hline
        Number of sampled individuals: &  & \\
         \hspace{0.5em} Urban & 20.0 (19.4) & 101.0 (75.6) \\
         \hspace{0.5em} Rural & 37.3 (31.9) & 186.3 (126.9) \\
         \hline
        Number of areas (out of 300): && \\
        \hspace{0.5em} With no sampled clusters & 7.7 (2.3) & 0.9 (0.6) \\
        \hspace{0.5em} With $< 5$ sampled clusters & 128.0 (5.8) & 4.3 (1.0) \\
    \end{tabular}
    \label{tab:sim_samplesizes}
\end{table}

\subsection{Candidate models \label{sec:candidate}}

For each dataset $f$ sampled from the simulated population, we use the design-based mean and variance estimates, $(\boldsymbol{\hat\theta}^{(f)},{\bf \hat V}^{(f)})$, to fit Fay-Herriot variance smoothing models, as defined in (\ref{eq:model_spec}), using the SRS, CSS, and CSM distributions defined in (\ref{eq:srs_sampling}), (\ref{eq:simple_approx}), and (\ref{eq:satt_approx}), respectively. For all models, $\boldsymbol{X}$ is the same matrix as in the data generating process, $\boldsymbol{\beta}$ is a vector of fixed effect regression coefficients, and $\boldsymbol{b}$ are BYM2 area-level random effects. For each of the three sampling distributions we use a structured and unstructured latent variance model, yielding six variance smoothing models which we will call `SRS-struct', `SRS-unstruct', `CSS-struct', `CSS-unstruct', `CSM-struct', and `CSM-unstruct'. For the structured latent variance model, $\boldsymbol{Z}$ is the same matrix as in the data generating process, $\boldsymbol{\eta}$ is a vector of fixed effect regression coefficients, and $\boldsymbol{g}$ are BYM2 area-level random effects. For the unstructured latent variance model, $\boldsymbol{Z}$ is a vector of $1$s, $\eta$ is an intercept term, and $\boldsymbol{g}$ is an IID normal area-level random effect. Note that whether the model is labeled structured or unstructured refers only to the latent variance model, as the latent mean model is structured in all variations. We choose these models to not only compare the performance of the sampling distributions, but to observe how these results may change if information regarding the structure of the variance is not available, as is often the case in applied settings. 

For comparison, we also fit two Fay-Herriot models which do not use variance smoothing. The first model, which we refer to as the `standard' model, is specified in (\ref{eq:latent_mean}). To compare performance when the design variance is correctly estimated without smoothing, which cannot be reliably done in a real data setting, we consider an `oracle' model which uses the sampling distribution, $$\hat\theta_i^{(f)}\mid \theta_i \sim\mathcal{N}\left(\theta_i,V _i\right),$$ where $V_i$, the empirical design variance, is calculated by taking the variance of the design-based estimates over $G=10000$ datasets sampled from the population surface using the stratified two-stage cluster sampling design described above. Specifically,
\begin{equation}
    V_i=\frac{1}{|\mathcal{G}_i|}\sum_{g\in \mathcal{G}_i}\left(\hat\theta_i^{(g)}-\frac{1}{|\mathcal{G}_i|}\sum_{g'\in \mathcal{G}_i}\hat\theta_i^{(g')}\right)^2
    \label{eq:emp_var}
\end{equation}
where $\mathcal{G}_i$ is the subset of dataset indices, $\mathcal{G}=(1,...,G)$, for which area $i$ has valid design-based mean and variance estimates (i.e., $m_{\cdot i}>1$). Note that the empirical design variance defined in (\ref{eq:emp_var}) is a Monte Carlo estimate of the true design variance \citep{sarndal}. In Appendix \ref{appendix:mc_error} of the online Supplemental Materials we obtain bootstrap estimates of the Monte Carlo error and show that this empirical estimate is a close approximation of the true design variance. Both comparison models use the same latent mean model as the four variance smoothing models. 

The same priors on regression parameters and hyperparameters are used for all models. All regression parameters in the latent mean model have a $\mathcal{N}(0,2^2)$ prior and all regression parameters in the latent variance model have a $\mathcal{N}(0,1^2)$ prior. In \ref{appendix:prior_sensitivity} of the online Supplemental Material we demonstrate that the models are fairly robust to the choice of priors on the regression parameters. For the PC priors on $\tau_b$ and $\tau_g$ we choose $u=1$ and $\alpha=0.01$. For $\phi_b$ and $\phi_g$ we choose a shrinkage prior of $\mathrm{Beta}(0.5,1)$ which has mean, $0.33$, and standard deviation, $0.3$. We use the \texttt{Stan} software (\citealp{standev}) for all models, and estimate the posterior distribution using 4 chains of 1000 iterations after 1000 iterations of burn-in. \texttt{Stan} diagnostics indicate convergence and efficient sampling for all models and simulations. For settings 1-4, the average run times across 100 simulations for the `SRS-struct', `SRS-unstruct', `CSS-struct', `CSS-unstruct', `CSM-struct', and `CSM-unstruct' models were 63, 46, 24, 23, 61, and 60 seconds, respectively. Due to increased sample size in setting 1a, models had longer run times: `SRS-struct', `SRS-unstruct', `CSS-struct', `CSS-unstruct', `CSM-struct', and `CSM-unstruct' models had average run times of 70, 53, 45, 44, 254, and 264 seconds, respectively.

\subsection{Results}

\subsubsection{Assessment of sampling distributions \label{sec:dist_comp}}

First we assess how closely the SRS, CSS, and CSM sampling distributions adhere to the empirical sampling distribution of the design variance estimator obtained via simulation. We do not use model estimates of $\boldsymbol{\sigma}^2$ and $\boldsymbol{\gamma}$, but simply plug in the correct values, which are defined in the simulation design, so that we may assess accuracy of these sampling distributions when parameters are correctly specified. For each area $i$, we want to evaluate the mean, variance, and skewness of $\hat V_i$ under each sampling distribution, $$\mu_{f,i}=\mathbb{E}_f[\hat V_i],\quad \sigma_{f,i}^2=\mathbb{E}_f\left[\left(\hat V_i-\mu_{f,i}\right)^2\right],\quad\mathrm{and}\quad\zeta_{f,i}=\mathbb{E}_f\left[\left(\frac{\hat V_i-\mu_{f,i}}{\sigma_{f,i}}\right)^3\right],$$
where $f$ is the empirical, SRS, CSS, or CSM sampling distribution. For the SRS, CSS, and CSM sampling distributions, we approximate these values by taking the sample averages,
$$\hat\mu_{f,i}=\frac{1}{L\cdot|\mathcal{G}_i|}\sum_{g\in\mathcal{G}_i}\sum_{\ell=1}^Lv_{f,i}^{(g,\ell)},\quad\hat\sigma^2_{f,i}=\frac{1}{L\cdot|\mathcal{G}_i|}\sum_{g\in\mathcal{G}_i}\sum_{\ell=1}^L\left(v_{f,i}^{(g,\ell)}-\hat\mu_{f,i}\right)^2,$$ $$\mathrm{and}\quad\hat\zeta_{f,i}=\frac{1}{L\cdot|\mathcal{G}_i|}\sum_{g\in\mathcal{G}_i}\sum_{\ell=1}^L\left(\frac{v_{f,i}^{(g,\ell)}-\hat\mu_{f,i}}{\hat\sigma_{f,i}}\right)^3,$$
where $L=10000$ and $v_{f,i}^{(g,\ell)}$ is the $\ell^{th}$ draw from sampling distribution $f$ under dataset $g$.

\begin{figure}[h!]
    \centering
    \includegraphics[width=0.9\linewidth]{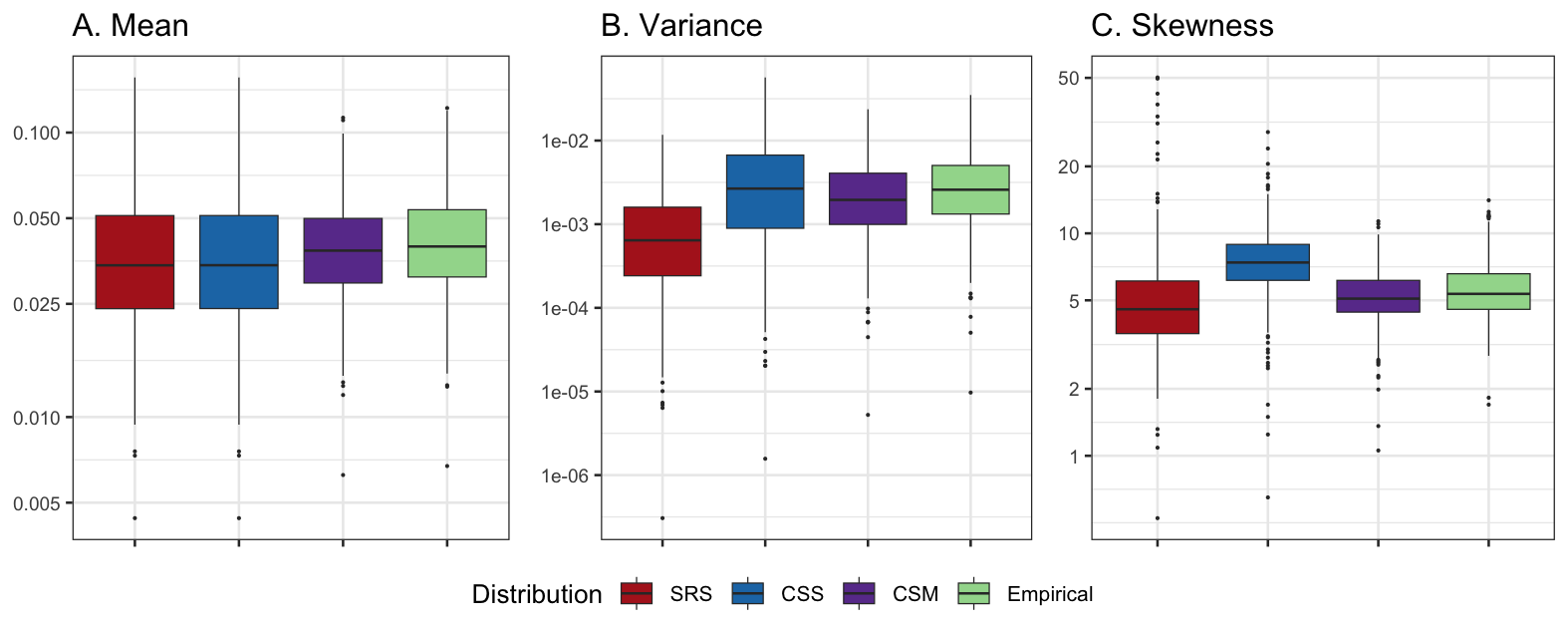}
    \caption{Mean, variance, and skewness of the design variance estimator under each sampling distribution for each area, under the first setting.}
    \label{fig:approx_moments}
\end{figure}

In Figure \ref{fig:approx_moments} we display the mean, variance, and skewness of the design variance estimator for each area under each sampling distribution, for the first simulation setting. The patterns observed here are similar across simulation settings, as shown in Appendix \ref{appendix:moments} of the online Supplemental Materials. In Figure \ref{fig:approx_moments}A, we observe that the mean under the CSM distribution is slightly closer to that of the empirical distribution, on average, while the CSS and SRS distributions slightly underestimate the mean, on average. This is a result of the CSS and SRS distributions assuming $\hat V_i$ is an unbiased estimator for $V_i$, which is not generally true for small sample sizes. In Figure \ref{fig:approx_moments}B, we observe that, in terms of variance, the CSS and CSM distributions are closer to the empirical distribution than the SRS distribution, which underestimates the variance as a result of not accounting for the additional variability introduced by using a complex survey design. In Figure \ref{fig:approx_moments}C, we observe that, in terms of skewness, the CSM distribution is closest to the empirical distribution, while the SRS and CSS distributions underestimate and overestimate, respectively. Note that skewness of a $\chi^2$ distribution is inversely proportional to its degrees of freedom, so this follows from the fact that the SRS distribution overstates the degrees of freedom, while the CSS understates the degrees of freedom, on average. 

Figure \ref{fig:approx_moments} verifies that a sampling distribution which accounts for the complex survey design is a better approximation of the empirical sampling distribution. The mean and variance of the design variance estimator under the CSS distribution, which only partially accounts for the survey design, is comparable to that under the CSM distribution, but the former distribution overstates skewness.

\subsubsection{Variance estimation \label{section:var_est_performance}}

\begin{figure}[h]
    \centering
    \includegraphics[width=1\linewidth]{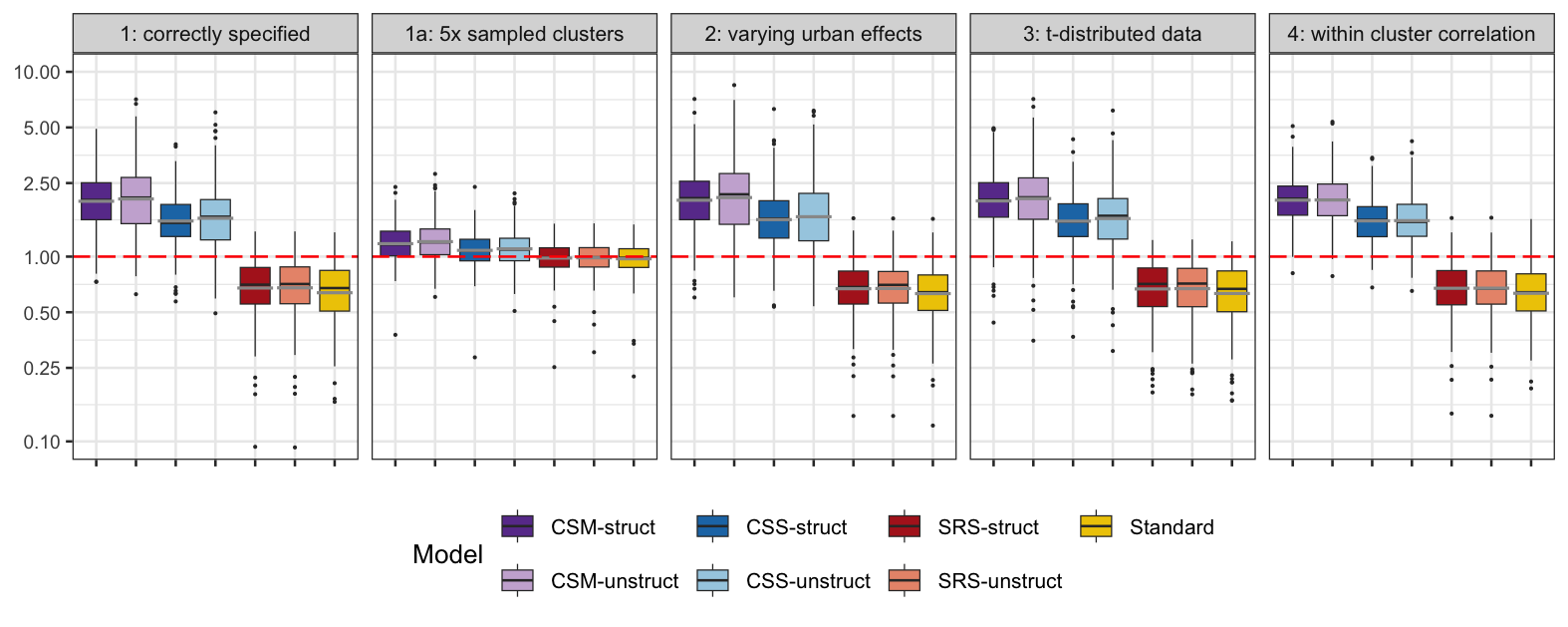}
    \caption{Boxplots of average estimate-to-truth ratios for the design variance for $300$ areas, for each model and setting. The average for each area is taken across 100 simulations.}
    \label{fig:sim_var}
\end{figure}

Now we evaluate how well the design variance is estimated for each area $i$, under each model. For each area $i$ and dataset $g$, the estimate-to-truth ratio for the design variance is defined by
\begin{equation}
    \frac{V^{(g)}_i(\widetilde\sigma^{2(g)}
_i)}{V_i}
\label{eq:est_to_truth_design}
\end{equation}

\noindent where $V_i$ is obtained from (\ref{eq:emp_var}), $V_i^{(g)}(\sigma^2)=\sigma^2/n_i^+$ for the SRS and CSS variance smoothing models, and $V_i^{(g)}(\sigma^2)=\sigma^2/n_i^\dagger$ for the CSM variance smoothing models. In Figure \ref{fig:sim_var} we present the average estimate-to-truth ratios of the design variance for each area, model, and setting, across $100$ simulations. We observe that, except for when sample size is sufficiently large, the CSS and CSM models overestimate the design variance, on average, while the SRS models underestimate the design variance. As expected, the design-based variance estimate, $\hat V_i$, underestimates the true design variance, except for when sample size is sufficiently large. As observed in Section \ref{sec:dist_comp}, the inflated degrees of freedom in the SRS distribution put too much confidence in $\hat V_i$, resulting in model-based estimates which are close to the downwardly biased design-based estimates. Conversely, the CSS and CSM distributions better account for the uncertainty of the empirical distribution of the design-based estimator, yielding more inflated variance estimates as a result of the noisy data.

\subsubsection{Mean estimation \label{section:mean_est_performance}}

\begin{figure}[h!]
     \centering
     \includegraphics[width=1\linewidth]{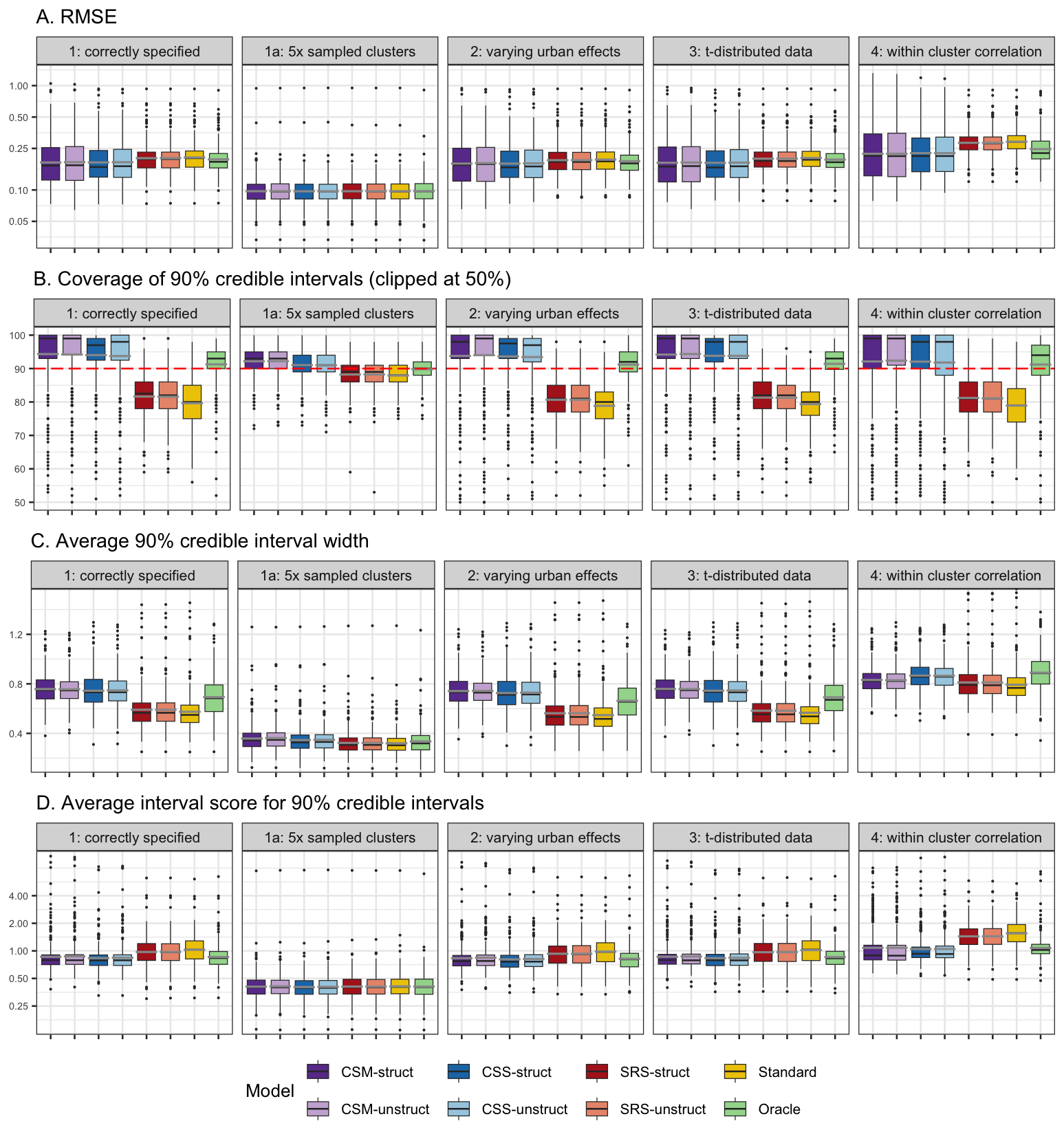}
      \caption{Boxplots of average RMSE of point estimates and coverage, average width, and average interval score of 90\% credible intervals for $300$ areas, for each model and setting. The average for each area is taken across $100$ simulations. The gray lines indicate the mean across areas and the red line in panel B indicates the nominal rate of the credible intervals.}
     \label{fig:sim_panel}
 \end{figure}

Now we compare the performance of the point estimates and credible intervals for the area-level means under each model using four metrics. Let $\tilde\theta_i^{(g)}$ denote the posterior mean estimate for $\theta_i$, for area $i$, using sample dataset $g$, and $\left(l_i^{(g)},u_i^{(g)}\right)$ denote its corresponding 90\% credible interval. First, we evaluate the root mean squared error (RMSE) for each area $i$, $$\sqrt{\frac{1}{100}\sum_{g=1}^{100}\left(\tilde\theta_i^{(g)}-\theta_i\right)^2}.$$ Next, we evaluate the coverage and average width of the 90\% credible intervals for each area $i$, $$\frac{1}{100}\sum_{g=1}^{100}I\left(\theta_i\in\left[l_i^{(g)},u_i^{(g)}\right]\right),\qquad\frac{1}{100}\sum_{g=1}^{100}\left(u_i^{(g)} - l_i^{(g)}\right).$$ Finally, we evaluate the average interval score, a proper scoring rule discussed in \cite{gneiting}, which favors narrow intervals containing the true parameter value. The average interval score for 90\% credible intervals for area $i$ is $$\frac{1}{100}\sum_{g=1}^{100}\left(u_i^{(g)} - l_i^{(g)}+\frac{2}{\alpha}\mathrm{max}\left\{0,l_i^{(g)}-\theta_i\right\}+\frac{2}{\alpha}\mathrm{max}\left\{0,\theta_i-u_i^{(g)}\right\}\right),$$ where we take $\alpha=0.1$. In Figure \ref{fig:sim_panel} we present the RMSE and the coverage, average width, and average interval score of 90\% credible intervals, per area, for each model, across $100$ sampled datasets for each setting.

In Figure \ref{fig:sim_panel}A we observe that the CSS and CSM variance smoothing models have similar or slightly lower RMSE, on average, compared to the SRS variance smoothing model and the standard model without variance smoothing. In Figures \ref{fig:sim_panel}B and \ref{fig:sim_panel}C we observe that, on average, the CSS and CSM variance smoothing models produce more conservative credible intervals and the SRS variance smoothing model and standard model produce credible intervals with undercoverage. This follows directly from the findings in Subsection \ref{section:var_est_performance} that the design variance is overestimated, on average, for the CSS and CSM variance smoothing models, and underestimated, on average, for the SRS variance smoothing model and standard model. We observe that when sample size increases, all credible intervals are closer to the nominal rate, on average. In Figure \ref{fig:sim_panel}D, we observe that the CSS and CSM variance smoothing models produce credible intervals which are more favorable, on average, than those under the SRS variance smoothing model and the standard model. Throughout, we observe that using a structured latent variance model has very little influence on model performance; the choice of sampling distribution for the design variance estimator is more influential. Moreover, using a sampling distribution which accounts for the complex design, whether it be partially (CSS) or fully (CSM), produces more favorable intervals estimates than the sampling distribution which does not account for the complex design (SRS).

\section{Data example: height-for-age z-score in Kenya \label{sec:data_example}}

We return to the example of HAZ at the Admin-2 level in the 2022 Kenya DHS, introduced in Section \ref{sec:introduction}. We obtain design-based estimates, $(\hat\theta_i,\hat V_i)$, for each Admin-2 area $i$ that has at least two sampled clusters ($288$ out of $300$ areas). We use $(\boldsymbol{\hat\theta},\boldsymbol{\hat V})$ to fit CSS and CSM variance smoothing Fay-Herriot models, as defined in Section \ref{sec:implementation},
\[
\begin{aligned}
\hat{\theta}_i \mid \theta_i, \sigma_i^2 
&\sim \mathcal{N}(\theta_i, V_i(\sigma_i^2)), 
&\quad
\hat{V}_i \mid \gamma, \sigma_i^2 
&\sim V_i(\sigma_i^2)\, \frac{c_i}{d_i(\gamma,\sigma_i^2)}\, \chi^2_{d_i(\gamma,\sigma_i^2)}, \\
\theta_i 
&= \boldsymbol{X}_i^\mathrm{T} \boldsymbol{\beta} + p_i \gamma + b_i, 
&\quad
\log(\sigma_i^2) 
&= \eta + g_i,
\end{aligned}
\]

\noindent where $\boldsymbol{X}_i$ is an area-specific auxiliary variable vector further defined below, $p_i$ is the area-specific urban under-5 population proportion, $(\boldsymbol{\beta},\gamma)$ is a vector of fixed effect regression coefficients, $\eta$ is an intercept term, and $\boldsymbol{b}$ and $\boldsymbol{g}$ are vectors of the residual between-area variation, modeled as area-level BYM2 random effects. The same priors and hyperpriors are used as in the simulation study. The constant $c_i$ and functions $V_i(\sigma_i^2)$ and $d_i(\gamma,\sigma^2_i)$, defined in (\ref{eq:simple_theoretical_var}) and (\ref{eq:sasw_theoretical_var}), are chosen to attain the CSS (\ref{eq:simple_approx}) or CSM (\ref{eq:satt_approx}) sampling distribution. For comparison, we also fit a standard Fay-Herriot model as defined in (\ref{eq:latent_mean}), where $\boldsymbol{X}_i$, $\boldsymbol{\beta}$, $\gamma$, and $\boldsymbol{b}$ retain the same definitions as in the variance smoothing models.

For the area-specific auxiliary variables in the latent mean model, in addition to urban under-5 population proportion, we use population density, yearly average temperature, yearly total precipitation, elevation, motorized travel time to healthcare, and yearly average nighttime lights. See \ref{appendix:kenya_covariates} of the online Supplemental Materials for more details. We do not use auxiliary variables for the latent variance model because, in practice, a principled approach for choosing covariates for the latent variance is less straightforward than for the latent mean, and in the simulation study we found their inclusion was not influential on model performance, even when informative covariates are chosen. We use the \texttt{Stan} software (\citealp{standev}) for all models, and estimate the posterior distribution using 4 chains of 1000 iterations after 1000 iterations of burn-in. The run times for the standard, CSS variance smoothing, and CSM variance smoothing models were 32, 29, and 69 seconds, respectively. \texttt{Stan} diagnostics indicate convergence and efficient sampling for all models. We present the parameter estimates under each model in \ref{app:parameter_estimates} of the online Supplemental Materials.

\begin{figure}[h!]
    \centering
    \includegraphics[width=0.8\linewidth]{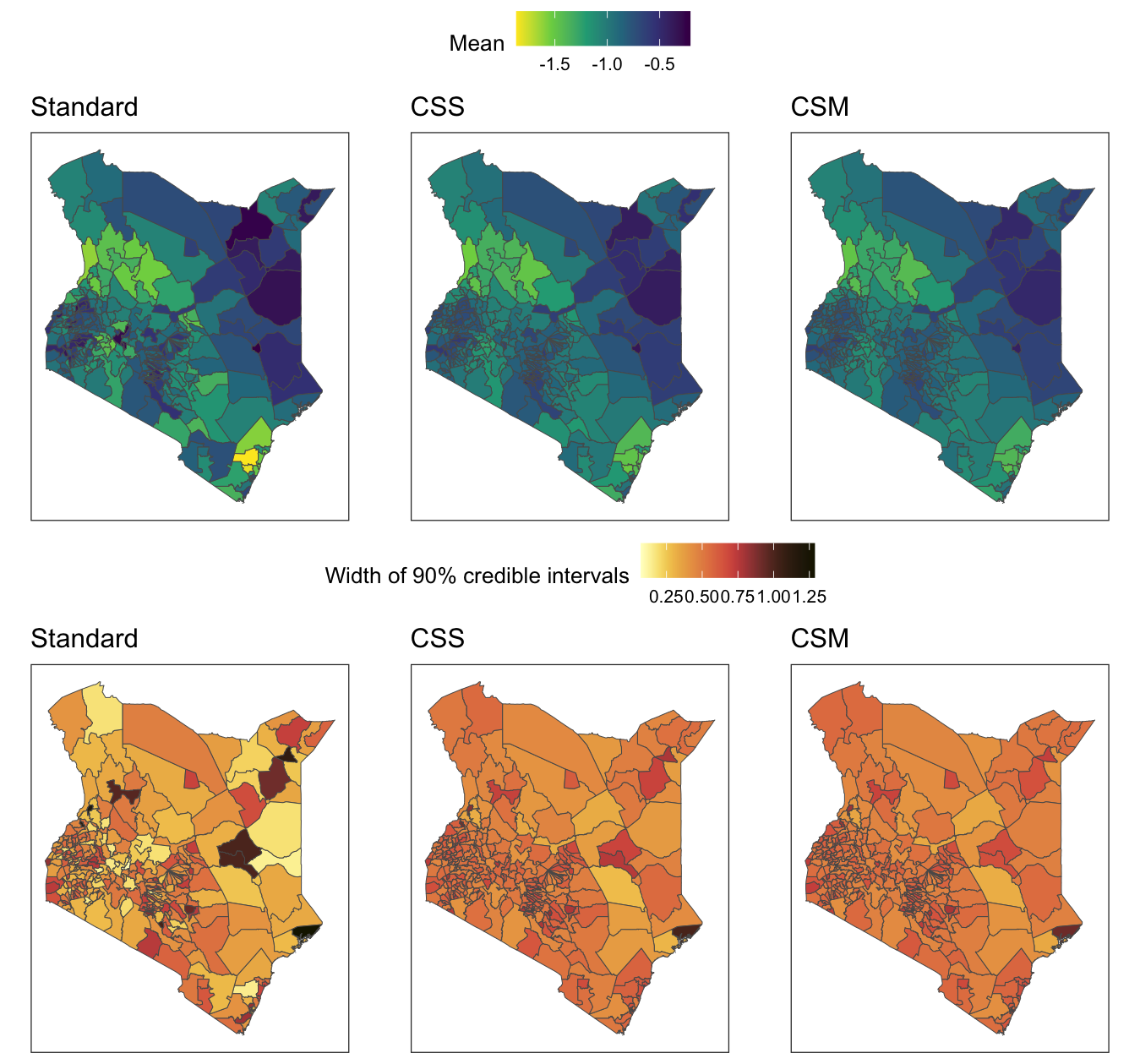}
    \caption{ Mean estimates and width of 90\% credible intervals for HAZ using 2022 Kenya DHS, under a standard Fay-Herriot model and CSS and CSM variance smoothing Fay-Herriot models.}
    \label{fig:kenya_est_map}
\end{figure}

\begin{figure}[h!]
    \centering
    \includegraphics[width=0.8\linewidth]{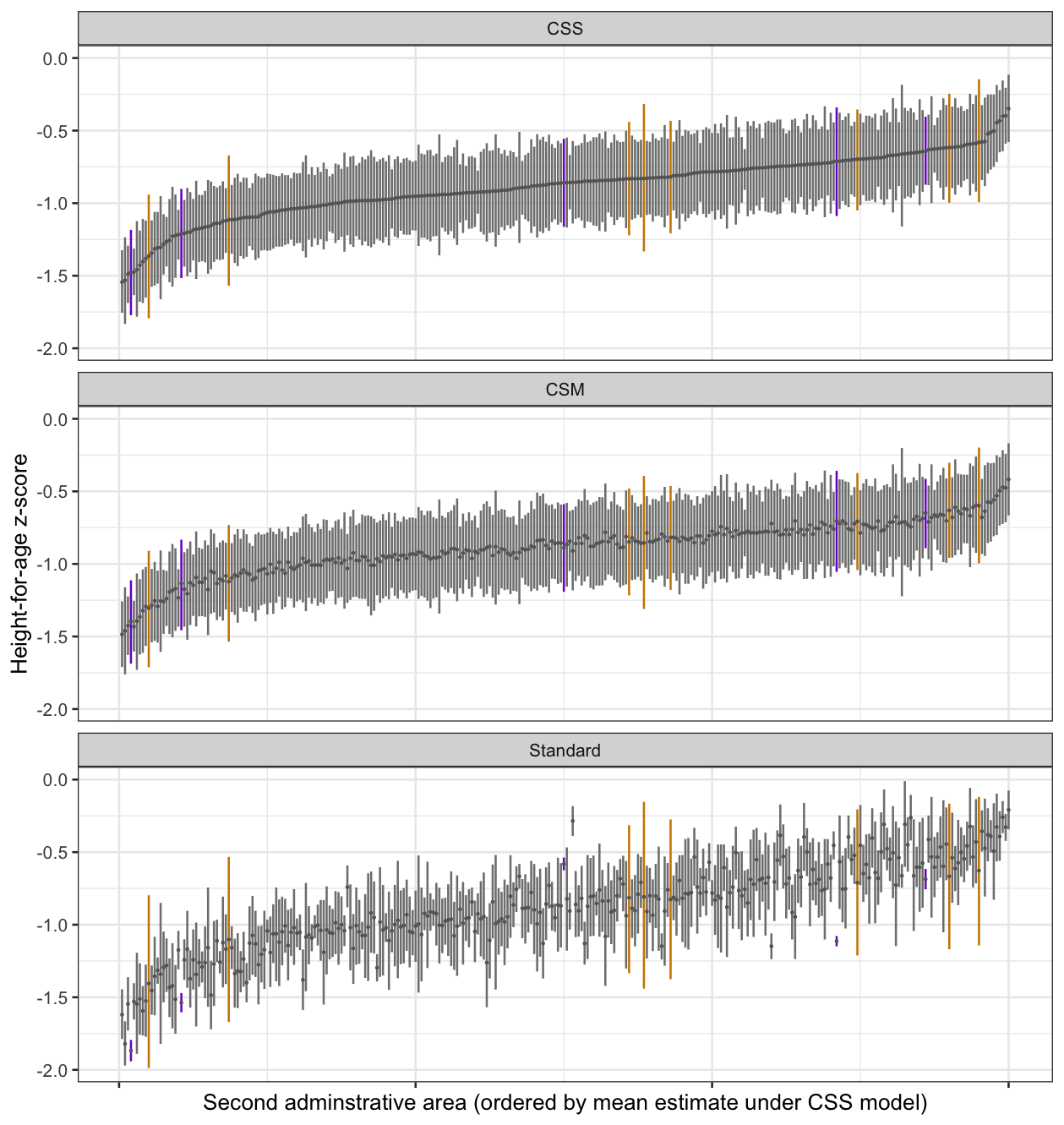}
    \caption{Mean estimates with 90\% credible intervals for HAZ  using 2022 Kenya DHS, under a standard Fay-Herriot model and CSS and CSM variance smoothing Fay-Herriot models. Admin-2 areas which have extremely narrow or wide credible intervals under the standard model are indicated in purple and orange, respectively.}
    \label{fig:caterpillar}
\end{figure}

Figure \ref{fig:kenya_est_map} presents the mean estimates and width of 90\% credible intervals under the standard model and each of the variance smoothing models. We observe that the mean estimates are fairly similar across models, with slightly more shrinkage towards the average in the variance smoothing models. The credible interval widths among the variance smoothing models are slightly higher on average than for the standard model, but there are several areas for which the standard model yields higher uncertainty than the variance smoothing models. Figure \ref{fig:caterpillar} more clearly illustrates the effect of the variance smoothing models on the credible intervals, where Admin-2 areas that have extremely narrow or wide intervals under the standard model are indicated in purple and orange, respectively. We observe that the variance smoothing models produce more reasonable uncertainty estimates. As observed in the simulation study, the CSS and CSM variance smoothing models perform similarly and, on average, produce more conservative intervals than the standard model. In \ref{app:kenya_scatter} we use scatter plots to compare the mean and standard deviation estimates under each model to the design-based estimates.

\begin{figure}[h!]
    \centering
    \includegraphics[width=0.8\linewidth]{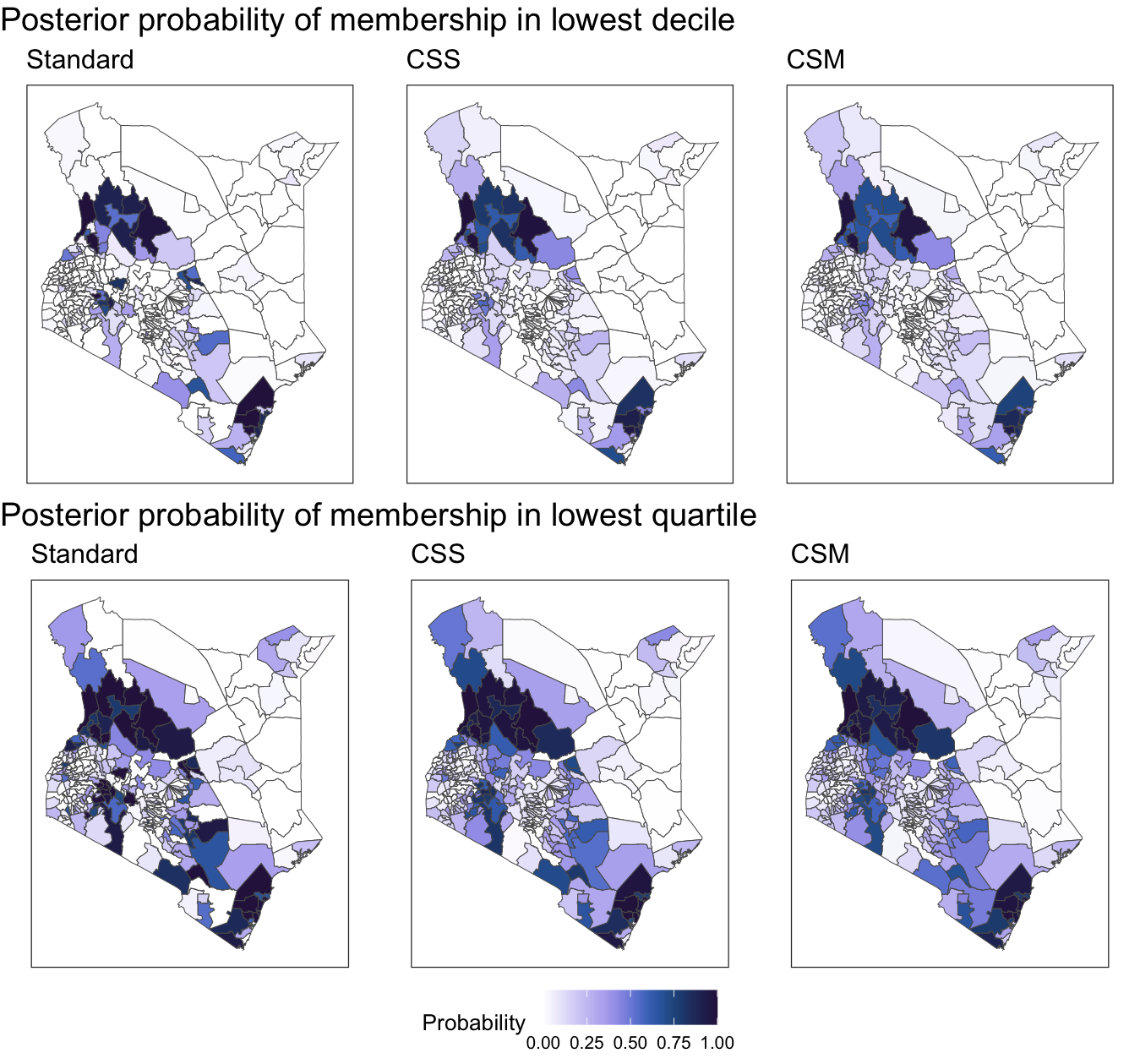}
    \caption{Posterior probabilities of membership in lowest decile and quartile for HAZ using 2022 Kenya DHS, under a standard Fay-Herriot model and CSS and CSM variance smoothing Fay-Herriot models.}
    \label{fig:kenya_postprob}
\end{figure}

In Figure \ref{fig:kenya_postprob} we visualize the uncertainty of estimates under each model by mapping the posterior probability that a given area has a HAZ which is in the lowest 10\% and 25\% of areas. The posterior probability that area $i$ is in the lowest $p\%$ of areas is calculated by, $$\frac{1}{4000}\sum_{s=1}^{4000}\mathbb{I}\left\{\frac{1}{K}\sum_{j=1}^K\mathbb{I}\left(\tilde\theta_i^{(s)}> \tilde\theta_j^{(s)}\right)\leq p\right\}$$ where $\tilde\theta_i^{(s)}$ and $\tilde\theta_j^{(s)}$ are the $s^{th}$ posterior draws for $\theta_i$ and $\theta_j$, respectively. We observe that the variance smoothing models produce very similar results to each other and, although uncertainty is higher on average for these models, the posterior probabilities are still quite informative in comparison to that of the standard model.

\section{Conclusion \label{sec:discussion}}

We began by establishing the limitations of assuming the design variance is known in a Fay-Herriot model, particularly when data is sparse. While GVF theory can be useful when estimating proportions or totals, their utility is limited in the context of small areas means because, generally, there is no clear choice of variance function. Alternatively, we consider jointly modeling the design variance along with the mean.  The sampling distributions used in existing joint smoothing Fay-Herriot models make strong implicit population assumptions which are often not stated explicitly and do not fully account for a complex survey design. This work thoroughly studies how various sampling distribution for the design variance estimator account for the survey design and how each of these models perform in a complex survey setting.

We considered three sampling distributions for the design variance estimator under superpopulation assumption (\ref{eq:pop_assumption}): the SRS distribution which follows from SRS; the CSS distribution which partially accounts for the complex survey design, arising from the common recommendation of using $\mathrm{df}=\# \mathrm{clusters} - \# \mathrm{strata}$; and the CSM distribution which fully accounts for the stratified cluster sampling survey design by allowing for unplanned domains and varying sampling weights and sample sizes. We illustrated through a simulation study that using a sampling distribution which accounts for the complex survey design (either partially or fully) is preferable to using a sampling distribution which ignores the survey design. We also found that model performance under these sampling distributions are robust to violations of several superpopulation assumptions, including departures from normality and presence of within-cluster correlation. We verified that the standard Fay-Herriot model can produce intervals with undercoverage in cases of sparse data and, although the CSS and CSM variance smoothing models produce more conservative intervals, they are preferable according to proper scoring rules.

We applied the CSS and CSM variance smoothing models to the estimation of HAZ using the 2022 Kenya DHS and showed that, while they produce more conservative intervals than the standard model, they can still provide informative posterior probabilities, which is particularly useful in policy and evaluation settings. From these results we conclude that both variance smoothing models which account for the complex survey design are preferred over the absence of variance smoothing, and the CSS variance smoothing model is a suitable choice because of its simplicity and more straightforward implementation relative to the CSM variance smoothing model. In Appendix J of the online Supplemental Materials we provide a vignette for implementing the CSS variance smoothing model.


While this work is motivated by data collected using a two-stage stratified cluster sampling design, the derivations and results hold generally for survey designs with cluster sampling where each individual in the cluster has the same sampling probability. These models could be extended for multi-stage sampling designs in which individuals in a cluster have different sampling probabilities, but further study is required to assess model performance in this setting. To extend these variance smoothing models for estimation of area-level proportions, we would need to assume that the distribution of the cluster-level proportions can be approximated by a normal distribution. If the conditions for approximating the binomial distribution of the cluster-level proportions with a normal distribution are met (i.e. non-rare events and sufficient sample size) these variance smoothing models may be a viable option. In the DHS setting, cluster sample sizes are generally too small for normality of proportions to be a reasonable assumption, however further study is required to determine how robust the model would be to violations of this assumption.  

\clearpage

\appendix

\bibliography{main}

\clearpage

\renewcommand{\thesection}{Appendix \Alph{section}}
\renewcommand{\thesubsection}{\Alph{section}\arabic{subsection}}
\setcounter{page}{1}
\setcounter{table}{0}
\setcounter{figure}{0}
\renewcommand{\thefigure}{S\arabic{figure}}
\renewcommand{\thetable}{S\arabic{table}}

\section{Form of the design variance estimator \label{app:var_form}}

\subsection{Decomposition of summations over clusters inside and outside the domain \label{app:var_decomp}}

Consider an area of interest, $i$, which is either a planned domain (i.e. $S_{hi}=S_h$ for all $h\in\mathcal{H}_i$) or an unplanned domain (i.e. $S_{hi}\varsubsetneq S_h$, for some $h\in\mathcal{H}_i$). For all $c\in S_h\setminus S_{hi}$ and $h\in\mathcal{H}_i$, let $w^\ast_c=0$, as these clusters are outside the area of interest. Following Equation 2.3-10 of \cite{korn_graubard}, we apply the Taylor linearization method to expression (\ref{eq:mean_estimator}) and obtain the design-consistent estimator for $V_i$ expressed in (\ref{eq:variance_longform}),
\[
    \hat V_i=\frac{1}{\left(\sum_{h\in\mathcal{H}_i}\sum_{c\in S_h}w^\ast _c\right)^2}\sum_{h\in\mathcal{H}_i}\frac{m_h}{m_h-1}\sum_{c\in S_h}\left[w^\ast _c(\bar y_c-\hat \theta_i)-\frac{1}{m_h}\sum_{c'\in S_h}w^\ast _{c'}(\bar y_{c'}-\hat\theta_i)\right]^2.
\]

To more clearly express the variance contributions of the clusters inside and outside of the area of interest, we can write (\ref{eq:variance_longform}) in terms of only its nonzero elements,
\begin{align*}
    \hat V_i &= \frac{1}{(\sum_{h\in\mathcal{H}_i}\sum_{c\in S_{hi}}w^\ast_c)^2}\sum_{h\in\mathcal{H}_i} \Biggl\{
    \begin{array}[t]{l}
        \dfrac{m_h}{m_h-1}\sum_{c\in S_{hi}}\left[w^\ast_c(\bar y_c-\hat\theta_i)-\dfrac{1}{m_h}\sum_{c'\in S_{hi}}w^\ast_{c'}(\bar y_{c'}-\hat\theta_i)\right]^2 \\[0.6em]
        + \dfrac{m_h}{m_h-1}\sum_{c\in S_h\setminus S_{hi}}\left[w_c^\ast(\bar y_c-\hat\theta_i)-\dfrac{1}{m_h}\sum_{c'\in S_{hi}}w^\ast_{c'}(\bar y_{c'}-\hat\theta_i)\right]^2
    \end{array}
    \Biggl\} \\[1.2em]
    &= \frac{1}{(\sum_{h\in\mathcal{H}_i}\sum_{c\in S_{hi}}w^\ast_c)^2}\sum_{h\in\mathcal{H}_i} \Biggl\{
    \begin{array}[t]{l}
        \dfrac{m_h}{m_h-1}\sum_{c\in S_{hi}}\left[w^\ast_c(\bar y_c-\hat\theta_i)-\dfrac{1}{m_h}\sum_{c'\in S_{hi}}w^\ast_{c'}(\bar y_{c'}-\hat\theta_i)\right]^2 \\[0.6em]
        + \dfrac{m_h}{m_h-1}\sum_{c\in S_h\setminus S_{hi}}\left[\dfrac{1}{m_h}\sum_{c'\in S_{hi}}w^\ast_{c'}(\bar y_{c'}-\hat\theta_i)\right]^2
    \end{array}
    \Biggl\} \\[1.2em]
    &= \frac{1}{(\sum_{h\in\mathcal{H}_i}\sum_{c\in S_{hi}}w^\ast_c)^2}\sum_{h\in\mathcal{H}_i} \Biggl\{
    \begin{array}[t]{l}
        \dfrac{m_h}{m_h-1}\sum_{c\in S_{hi}}\left[w^\ast_c(\bar y_c-\hat\theta_i)-\dfrac{1}{m_h}\sum_{c'\in S_{hi}}w^\ast_{c'}(\bar y_{c'}-\hat\theta_i)\right]^2 \\[0.6em]
        + \dfrac{m_h-m_{hi}}{(m_h-1)m_h}\left[\sum_{c'\in S_{hi}}w^\ast_{c'}(\bar y_{c'}-\hat\theta_i)\right]^2,
    \end{array}
    \Biggl\}
\end{align*}

\noindent where the first term inside the larger summation is a sum over the clusters inside the area of interest and the second term is a sum over the clusters outside the area of interest. From this expression we observe that when the area of interest is an unplanned domain, the number of clusters in the larger planned domain is incorporated in the last term, thus accounting for the additional variability introduced by the fact that the number of sampled clusters in the unplanned domain is random. Moreover, if the area of interest is a planned domain ($m_h=m_{hi}$), the last term becomes zero because the number of sampled clusters is fixed, and the estimator simplifies to
\[
    \hat V_i=\frac{1}{\left(\sum_{h\in\mathcal{H}_i}\sum_{c\in S_{hi}}w^\ast _c\right)^2}\sum_{h\in\mathcal{H}_i}\frac{m_{hi}}{m_{hi}-1}\sum_{c\in S_{hi}}\left[w^\ast _c(\bar y_c-\hat \theta_i)-\frac{1}{m_{hi}}\sum_{c'\in S_{hi}}w^\ast _{c'}(\bar y_{c'}-\hat\theta_i)\right]^2.
\]

\subsection{Form of the variance estimator under CSS distribution assumptions (Equation \ref{eq:variance_simple}) \label{app:simple_form}}

In this section we show that the design variance estimator presented in (\ref{eq:variance_longform}),
\[
    \hat V_i=\frac{1}{\left(\sum_{h\in\mathcal{H}_i}\sum_{c\in S_h}w^\ast _c\right)^2}\sum_{h\in\mathcal{H}_i}\frac{m_h}{m_h-1}\sum_{c\in S_h}\left[w^\ast _c(\bar y_c-\hat \theta_i)-\frac{1}{m_h}\sum_{c'\in S_h}w^\ast _{c'}(\bar y_{c'}-\hat\theta_i)\right]^2,
\]

\noindent simplifies to the estimator presented in (\ref{eq:variance_simple}),
\[
\hat V_i=\frac{1}{m_{\cdot i}(m_{\cdot i}-|\mathcal{H}_i|)}\sum_{h\in\mathcal{H}_i}\sum_{c\in S_{hi}}\left[\bar y_c-\frac{1}{m_{hi}}\sum_{c'\in S_{hi}}\bar y_{c'}\right]^2,
\]

\noindent under the survey design properties presented in Subsection \ref{section:simple}.

Under assumption (i), the area of interest is a planned domain, so (\ref{eq:variance_longform}) can be written as
\[
    \hat V_i=\frac{1}{\left(\sum_{h\in\mathcal{H}_i}\sum_{c\in S_{hi}}w^\ast _c\right)^2}\sum_{h\in\mathcal{H}_i}\frac{m_{hi}}{m_{hi}-1}\sum_{c\in S_{hi}}\left[w^\ast _c(\bar y_c-\hat \theta_i)-\frac{1}{m_{hi}}\sum_{c'\in S_{hi}}w^\ast _{c'}(\bar y_{c'}-\hat\theta_i)\right]^2.
\]

Then under assumption (ii), the same number of clusters are sampled from each stratum in area $i$ (i.e., $m_{hi}=m_{\cdot i}/|\mathcal{H}_i|$) so it follows
\[
\hat V_i=\frac{1}{\left(\sum_{h\in\mathcal{H}_i}\sum_{c\in S_{hi}}w^\ast _c\right)^2}\frac{m_{\cdot i}}{m_{\cdot i}-|\mathcal{H}_i|}\sum_{h\in\mathcal{H}_i}\sum_{c\in S_{hi}}\left[w^\ast _c(\bar y_c-\hat \theta_i)-\frac{1}{m_{hi}}\sum_{c'\in S_{hi}}w^\ast _{c'}(\bar y_{c'}-\hat\theta_i)\right]^2.
\]

Then under assumptions (iii) and (iv), $w_c^\ast=w^\ast_0$ for all $c\in\cup_{h\in\mathcal{H}_i}S_{hi}$, so it follows,
\[
\begin{aligned}
\hat V_i&=\frac{w^{\ast 2}_0}{\left(m_{\cdot i}w^\ast_0 \right)^2}\frac{m_{\cdot i}}{m_{\cdot i}-|\mathcal{H}_i|}\sum_{h\in\mathcal{H}_i}\sum_{c\in S_{hi}}\left[\left(\bar y_c-\hat \theta_i\right)-\frac{1}{m_{hi}}\sum_{c'\in S_{hi}}\left(\bar y_{c'}-\hat\theta_i\right)\right]^2 \\
& =\frac{1}{m_{\cdot i}^2}\frac{m_{\cdot i}}{m_{\cdot i}-|\mathcal{H}_i|}\sum_{h\in\mathcal{H}_i}\sum_{c\in S_{hi}}\left[\left(\bar y_c-\frac{1}{m_{hi}}\sum_{c'\in S_{hi}}\bar y_{c'}\right) -\left(1-\frac{m_{hi}}{m_{hi}}\right)\hat \theta_i\right]^2 \\ 
& =\frac{1}{m_{\cdot i}(m_{\cdot i}-|\mathcal{H}_i|)}\sum_{h\in\mathcal{H}_i}\sum_{c\in S_{hi}}\left[\bar y_c-\frac{1}{m_{hi}}\sum_{c'\in S_{hi}}\bar y_{c'}\right]^2.
\end{aligned}
\]

\subsection{Matrix form of the variance estimator (Equation \ref{eq:matrixform}) \label{matrix_form_appendix}}

In this section we show that the variance estimator (\ref{eq:variance_longform}), $$\hat V_i=\frac{1}{\left(\sum_{h\in \mathcal{H}_i}\sum_{c\in S_h}w^\ast _c\right)^2}\sum_{h\in \mathcal{H}_i}\frac{m_h}{m_h-1}\sum_{c\in S_h}\left[w^\ast _c(\bar y_c-\hat \theta_i)-\frac{1}{m_h}\sum_{c'\in S_h}w^\ast _{c'}(\bar y_{c'}-\hat\theta_i)\right]^2$$ can be written in the form of expression (\ref{eq:matrixform}), $$\hat V_i=\frac{1}{\left({\bf 1}^\mathrm{T}\boldsymbol{w_i}^\ast \right)^2}\boldsymbol{\bar y}_i^\mathrm{T}{\bf M}_{\boldsymbol{i}}\boldsymbol{\bar y}_i.$$

Recall from Subsection \ref{satt_section}, $\boldsymbol{\bar y_i}$ is composed of subvectors, $(\boldsymbol{\bar y}_{\boldsymbol{i},h})_{h\in\mathcal{H}_i}$, with elements $(\bar y_c)_{c\in S_h}$ and ${\bf M}_{\boldsymbol{i}}={\bf T}_{\boldsymbol{i}}^\mathrm{T}{\bf B}_{\boldsymbol{i}}{\bf T}_{\boldsymbol{i}}$. Let ${\bf B}_{\boldsymbol{i}}$ be a block diagonal matrix with blocks $\left({\bf B}_h\right)_{h\in\mathcal{H}_i}$ such that ${\bf B}_h=\frac{m_h}{m_h-1}\left({\bf I}_{m_h}-\frac{1}{m_h}{\bf 1}_{m_h}{\bf 1}^\mathrm{T}_{m_h}\right)$. Let $\boldsymbol{w_i^\ast}$ be a vector composed of subvectors, $(\boldsymbol{w}^\ast_h)_{h\in\mathcal{H}_i}$, with elements $\{w^\ast _c\}_{c\in S_h}$ and define ${\bf T}_{\boldsymbol{i}}={\bf W}_{\boldsymbol{i}}\left({\bf I}_{m_{\cdot i}}-\frac{{\bf 1}\boldsymbol{w_i}^{\ast \mathrm{T}}}{{\bf 1}^\mathrm{T}\boldsymbol{w_i}^\ast }\right)$, where ${\bf W}_{\boldsymbol{i}}=\mbox{diag}\left(\boldsymbol{w_i}^\ast \right)$. Note that $(\boldsymbol{\bar y}_i,{\bf B}_{\boldsymbol{i}},\boldsymbol{w_i^\ast})$ must maintain consistent ordering. By definition, ${\bf 1}^\mathrm{T}\boldsymbol{w_i}^\ast =\sum_{h\in\mathcal{H}_i}\sum_{c\in S_h}w^\ast _c$, so to prove equality it suffices to show
\begin{equation}
    \left({\bf T}_{\boldsymbol{i}}\boldsymbol{\bar y_i}\right)^\mathrm{T}{\bf B}_{\boldsymbol{i}}{\bf T}_{\boldsymbol{i}}\boldsymbol{\bar y_i}=\sum_{h\in\mathcal{H}_i}\frac{m_h}{m_h-1}\sum_{c\in S_h}\left[w^\ast _c(\bar y_c-\hat \theta_i)-\frac{1}{m_h}\sum_{c'\in S_h}w^\ast _{c'}(\bar y_{c'}-\hat\theta_i)\right]^2.
    \label{eq:target_equality}
\end{equation}

By definition, ${\bf T}_{\boldsymbol{i}}$ is a square matrix of dimension $\sum_{h\in\mathcal{H}_i}m_h$ with diagonal elements, ${\bf T}_{\boldsymbol{i},cc}=w^\ast _c-\frac{w_c^{\ast 2}}{{\bf 1}^\mathrm{T}\boldsymbol{w_i}^\ast }$, and off-diagonal elements, ${\bf T}_{\boldsymbol{i},cc'}=-\frac{w^\ast _cw^\ast _{c'}}{{\bf 1}^\mathrm{T}\boldsymbol{w_i}^\ast }$ for $(c,c')\in S=\cup_{h\in\mathcal{H}_i}S_h$. Then it follows that ${\bf T}_{\boldsymbol{i}}\boldsymbol{\bar y_i}$ is a vector composed of subvectors, $(\boldsymbol{L}_h)_{h\in\mathcal{H}_i}$, where $\boldsymbol{L}_h$ is a vector of length $m_h$ with elements $$\boldsymbol{L}_{h,c}=w^\ast _c\bar y_c-w^\ast _c\sum_{c'\in S}\frac{w^\ast _{c'}\bar y_{c'}}{{\bf 1}^\mathrm{T}\boldsymbol{w_i}^\ast }=w^\ast _c(\bar y_c-\hat\theta_i)$$ for $c\in S_h$, and \[
\begin{aligned}
        ({\bf T}_{\boldsymbol{i}}\boldsymbol{\bar y_i})^\mathrm{T}{\bf B}_{\boldsymbol{i}}{\bf T}_{\boldsymbol{i}}\boldsymbol{\bar y_i} & =\sum_{h\in\mathcal{H}_i} \boldsymbol{L}_h^\mathrm{T}{\bf B}_h\boldsymbol{L}_h \\
        & =\sum_{h\in\mathcal{H}_i}\frac{m_h}{m_h-1}\boldsymbol{L}_h^\mathrm{T}\left({\bf I}_{m_h}-\frac{1}{m_h}{\bf 1}_{m_h}{\bf 1}^\mathrm{T}_{m_h}\right)\boldsymbol{L}_h.
\end{aligned}\]

Observe that the matrix $\left({\bf I}_{m_h}-\frac{1}{m_h}{\bf 1}_{m_h}{\bf 1}^\mathrm{T}_{m_h}\right)$ is idempotent, so it is equal to the square of itself, and $\left({\bf I}_{m_h}-\frac{1}{m_h}{\bf 1}_{m_h}{\bf 1}^\mathrm{T}_{m_h}\right)\boldsymbol{L}_h$ is an $m_h-$length vector with elements

$$\left[\left({\bf I}_{m_h}-\frac{1}{m_h}{\bf 1}_{m_h}{\bf 1}^\mathrm{T}_{m_h}\right)\boldsymbol{L}_h\right]_c=\boldsymbol{L}_{h,c}-\frac{1}{m_h}\sum_{c'\in S_h}\boldsymbol{L}_{h,c'}$$ for $c\in S_h$, so it follows that

$$({\bf T}_{\boldsymbol{i}}\boldsymbol{\bar y_i})^\mathrm{T}{\bf B}_{\boldsymbol{i}}{\bf T}_{\boldsymbol{i}}\boldsymbol{\bar y_i}=\sum_{h\in\mathcal{H}_i}\frac{m_h}{m_h-1}\sum_{c\in S_h}\left(\boldsymbol{L}_{h,c}-\frac{1}{m_h}\sum_{c'\in S_h}\boldsymbol{L}_{h,c'}\right)^2$$

\noindent which results in the equality in expression $(\ref{eq:target_equality})$.

\clearpage

\setcounter{page}{1}
\section{Accuracy of Satterthwhaite approximation \label{appendix:satt_approx}}

In this section we evaluate how well the Satterthwhaite approximation (\ref{eq:satt_approx}) approximates the CSM distribution (\ref{eq:exact_dist}). We find that the approximation is highly accurate using sampling weights and sample sizes from various Admin-2 areas in the Kenya 2022 DHS and different sets of parameters.

\begin{figure}[h!]
    \centering
    \includegraphics[width=0.85\linewidth]{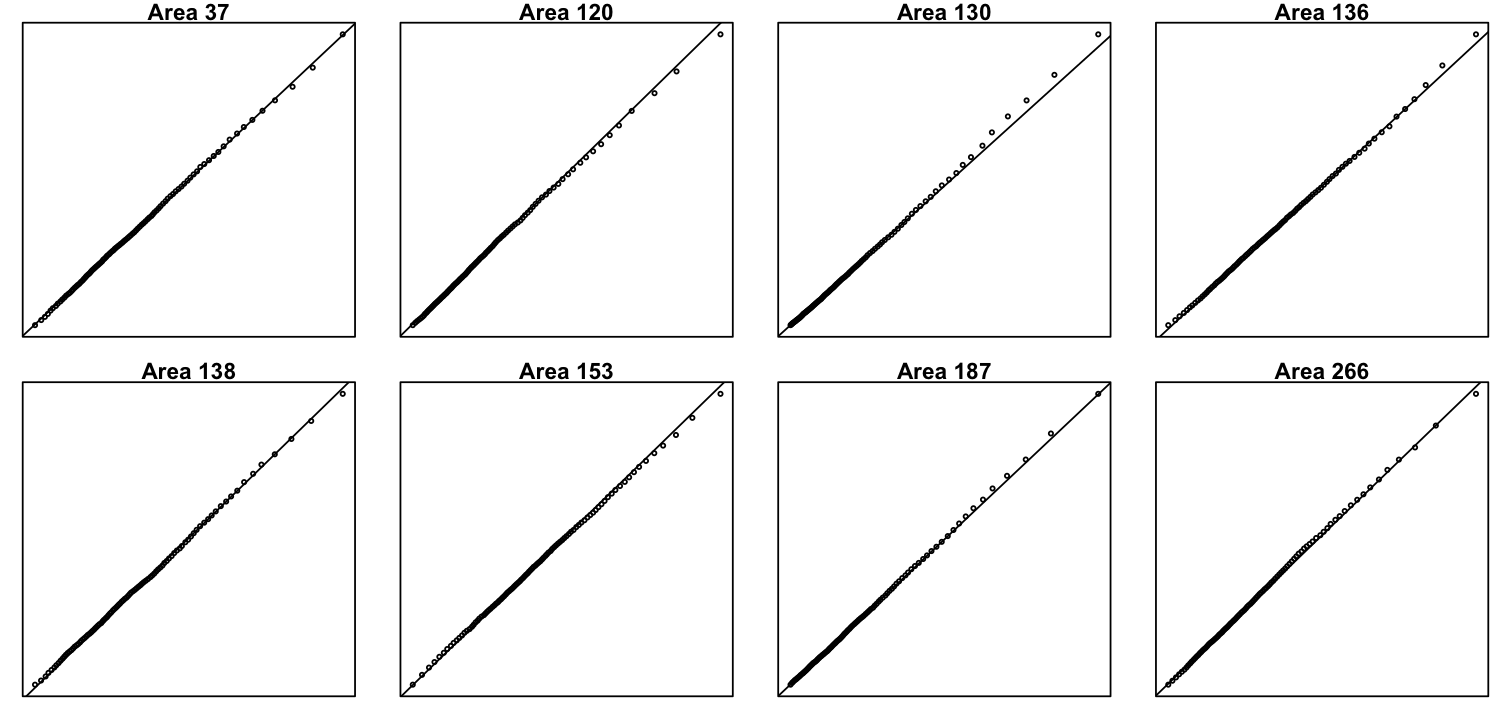}
    \caption{Quantile-quantile plots comparing CSM distribution, on the x-axis, to its Satterthwhaite approximation, on the y-axis, when $\gamma=1$ and $\sigma^2=1$, using the sampling weights and sample sizes from 8 Admin-2 areas in the 2022 Kenya DHS.}
\end{figure}

\begin{figure}[h!]
    \centering
    \includegraphics[width=0.85\linewidth]{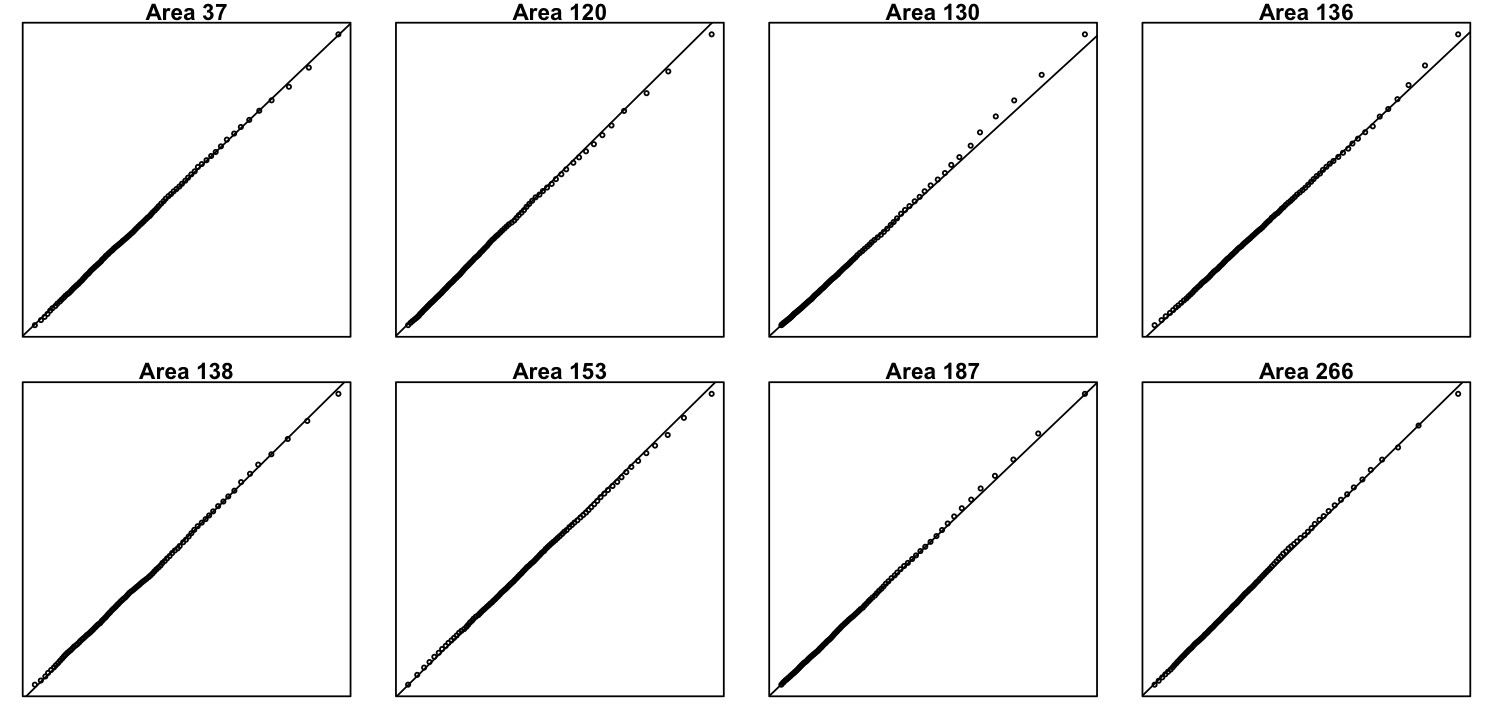}
    \caption{Quantile-quantile plots comparing CSM distribution, on the x-axis, to its Satterthwhaite approximation, on the y-axis, when $\gamma=2$ and $\sigma^2=4$, using the sampling weights and sample sizes from 8 Admin-2 areas in the 2022 Kenya DHS.}
\end{figure}

\clearpage

\setcounter{page}{1}
\section{Bias of design variance estimator under CSM sampling distribution (Equation \ref{eq:satt_approx})\label{appendix:bias_derivation}}

Recall from Subsection \ref{satt_section}, ${\bf D}_{\boldsymbol{i}}$ is a a diagonal matrix with elements $(1/n_c)_{c\in \cup_{h\in\mathcal{H}_i}S_h}$, $\boldsymbol{w_i}^\ast$ is a vector composed of subvectors, $(\boldsymbol{w}^\ast_h)_{h\in\mathcal{H}_i}$, with elements $\left(w^\ast _c\right)_{c\in S_h}$, and ${\bf M}_{\boldsymbol{i}}={\bf T}_{\boldsymbol{i}}^\mathrm{T}{\bf B}_{\boldsymbol{i}}{\bf T}_{\boldsymbol{i}}$ where ${\bf T}_{\boldsymbol{i}}=\mbox{diag}\left(\boldsymbol{w_i}^\ast \right)\left({\bf I}_{m_{\cdot i}}-\frac{{\bf 1}\boldsymbol{w_i}^{\ast \mathrm{T}}}{{\bf 1}^\mathrm{T}\boldsymbol{w_i}^\ast }\right)$ and ${\bf B}_{\boldsymbol{i}}$ is a block diagonal matrix with blocks $\left({\bf B}_h\right)_{h\in\mathcal{H}_i}$ such that ${\bf B}_h=\frac{m_h}{m_h-1}\left({\bf I}_{m_h}-\frac{1}{m_h}{\bf 1}_{m_h}{\bf 1}^\mathrm{T}_{m_h}\right)$. Also recall, $\boldsymbol{q_i}$ is the vector of non-zero eigenvalues of ${\bf D}_{\boldsymbol{i}}^{1/2}{\bf M}_{\boldsymbol{i}}{\bf D}_{\boldsymbol{i}}^{1/2}$ and $\delta_{ij}=\frac{1}{\sigma_i^2}\left(\boldsymbol{v}_{\boldsymbol{i}j}^\mathrm{T}\boldsymbol{\gamma_i}\right)^2 $, where ${\boldsymbol{v}_{\boldsymbol{i}j}}$ denotes the $m_{\cdot i}$-length eigenvector corresponding to eigenvalue $q_{ij}$, scaled such that $\boldsymbol{v}_{\boldsymbol{i}j}^\mathrm{T}{\bf D}_{\boldsymbol{i}}\boldsymbol{v}_{\boldsymbol{i}j}=1$ and $\boldsymbol{\gamma_i}$ is a vector composed of subvectors, $\left(\gamma_h\boldsymbol{1}_{m_h}\right)_{h\in\mathcal{H}_i}$. The CSM distribution (\ref{eq:satt_approx}) is defined as
\[
\hat V_i\mid \boldsymbol{\gamma_i},\sigma^2_i \sim \frac{\sigma^2_i/n_i^\dagger}{\boldsymbol{w_i}^{\ast \mathrm{T}}{\bf D}_{\boldsymbol{i}}\boldsymbol{w_i}^{\ast }}\frac{Q_2}{2Q_1}\chi^2_{\frac{2Q_1^2}{Q_2}},
\]

\noindent where $n_i^\dagger= \frac{(\sum_{h\in\mathcal{H}_i}\sum_{c\in S_{hi}}w^\ast_c)^2}{\sum_{h\in\mathcal{H}_i}\sum_{c\in S_{hi}}w^{\ast2}_c/n_c}$, $Q_1=\sum_j^{r_i} q_{ij}(1+\delta_{ij})$ and $Q_2=2\sum_j^{r_i} q_{ij}^2(1+2\delta_{ij})$.

Then it follows that \[\mathbb{E}[\hat V_i\mid \boldsymbol{\gamma_i},\sigma^2_i]=\frac{Q_1}{\boldsymbol{w_i}^{\ast \mathrm{T}}{\bf D}_{\boldsymbol{i}}\boldsymbol{w_i}^{\ast }}\frac{\sigma^2_i}{n_i^\dagger}\]

\noindent so $\hat V_i\mid \sigma^2_i,\boldsymbol{\gamma_i}$ is a biased estimator for the theoretical design variance, $\sigma^2_i/n_i^\dagger$, by a factor of $Q_1/(\boldsymbol{w_i}^{\ast \mathrm{T}}{\bf D}_{\boldsymbol{i}}\boldsymbol{w_i}^{\ast })$. We will provide conditions under which $Q_1/(\boldsymbol{w_i}^{\ast \mathrm{T}}{\bf D}_{\boldsymbol{i}}\boldsymbol{w_i}^{\ast })<1$.

From the definitions of $\boldsymbol{q_i}$ and $\boldsymbol{\delta_i}$ it follows that \[Q_1=\sum_jq_{ij} + \sum_jq_{ij}\delta_{ij}=\mathrm{tr}({\bf M}_{\boldsymbol{i}}{\bf D}_{\boldsymbol{i}}) + \frac{1}{\sigma_i^2}\boldsymbol{\gamma_i}^\mathrm{T}{\bf M}_{\boldsymbol{i}}{\bf D}_{\boldsymbol{i}}{\bf M}_{\boldsymbol{i}}\boldsymbol{\gamma_i}.\]

\noindent The first term,  $\mathrm{tr}({\bf M}_{\boldsymbol{i}}{\bf D}_{\boldsymbol{i}})$, can be expressed as $\boldsymbol{w_i}^{\ast \mathrm{T}}{\bf D}_{\boldsymbol{i}}\boldsymbol{w_i}^{\ast } - R_i$, where

\begin{align*}
R_i=\sum_{h\in\mathcal{H}_i}\frac{m_h}{m_h-1}\sum_{c\in S_h} \Bigg\{& \left[\frac{w_c^\ast}{\sqrt{n_c}}-\frac{1}{m_h}\sum_{c'\in S_h}\frac{w_{c'}^\ast}{\sqrt{n_{c'}}}\right]\\
& -\frac{\sum_{h\in\mathcal{H}_i}\sum_{c'\in S_h}w_{c'}^\ast/n_{c'}}{\sum_{h\in\mathcal{H}_i}\sum_{c'\in S_h}w_{c'}^\ast}\left[w_c^\ast-\frac{1}{m_h}\sum_{c'\in S_h}w_{c'}^\ast\right]^2\Bigg\},
\end{align*}

\noindent so $$\frac{Q_1}{\boldsymbol{w_i}^{\ast \mathrm{T}}{\bf D}_{\boldsymbol{i}}\boldsymbol{w_i}^{\ast }}
=\frac{\boldsymbol{w_i}^{\ast \mathrm{T}}{\bf D}_{\boldsymbol{i}}\boldsymbol{w_i}^{\ast } - R_i + \frac{1}{\sigma_i^2}\boldsymbol{\gamma_i}^\mathrm{T}{\bf M}_{\boldsymbol{i}}{\bf D}_{\boldsymbol{i}}{\bf M}_{\boldsymbol{i}}\boldsymbol{\gamma_i}}{\boldsymbol{w_i}^{\ast \mathrm{T}}{\bf D}_{\boldsymbol{i}}\boldsymbol{w_i}^{\ast }}=1+\frac{\frac{1}{\sigma_i^2}\boldsymbol{\gamma_i}^\mathrm{T}{\bf M}_{\boldsymbol{i}}{\bf D}_{\boldsymbol{i}}{\bf M}_{\boldsymbol{i}}\boldsymbol{\gamma_i}-R_i}{\boldsymbol{w_i}^{\ast \mathrm{T}}{\bf D}_{\boldsymbol{i}}\boldsymbol{w_i}^{\ast }}.$$ Then it follows that $Q_1/(\boldsymbol{w_i}^{\ast \mathrm{T}}{\bf D}_{\boldsymbol{i}}\boldsymbol{w_i}^{\ast })<1$ when $R_i>\boldsymbol{\gamma_i}^\mathrm{T}{\bf M}_{\boldsymbol{i}}{\bf D}_{\boldsymbol{i}}{\bf M}_{\boldsymbol{i}}\boldsymbol{\gamma_i}/\sigma_i^2$. Note that both $R_i$ and $\boldsymbol{\gamma_i}^\mathrm{T}{\bf M}_{\boldsymbol{i}}{\bf D}_{\boldsymbol{i}}{\bf M}_{\boldsymbol{i}}\boldsymbol{\gamma_i}/\sigma_i^2$ are nonnegative. The second term, $\boldsymbol{\gamma_i}^\mathrm{T}{\bf M}_{\boldsymbol{i}}{\bf D}_{\boldsymbol{i}}{\bf M}_{\boldsymbol{i}}\boldsymbol{\gamma_i}/\sigma_i^2$, can be expressed as \[\left(\frac{m_h}{m_h-1}\right)^2\left(\frac{\gamma_h-\bar\gamma^W}{\sigma_i}\right)^2\sum_{c\in S_h}\frac{1}{n_c}\left(w_c^\ast-\frac{1}{m_h}\sum_{c'\in S_h}w_{c'}^\ast\right)^2\]

\noindent where $\bar\gamma^W=\frac{\sum_h\gamma_h\sum_{c\in S_h}w_c^\ast}{\sum_h\sum_{c\in S_h}w_c^\ast}$. We observe that $\boldsymbol{\gamma_i}^\mathrm{T}{\bf M}_{\boldsymbol{i}}{\bf D}_{\boldsymbol{i}}{\bf M}_{\boldsymbol{i}}\boldsymbol{\gamma_i}/\sigma_i^2$ tends to zero when $w_c^\ast$ takes a similar value for all $c\in S_h$, which is generally the case unless there are extreme differences at the cluster size re-enumeration stage. This term also decreases when the within-stratum variance is larger than the between-stratum variance. On the other hand, $R_i$ only tends to zero when, in addition to $w_c^\ast$, $n_c$ also takes similar values for all $c\in S_h$, which is a less reasonable assumption (see \ref{sample_size_appendix} of the online Supplemental Materials).

\clearpage

\setcounter{page}{1}
\section{Additional simulation results}

\subsection{Monte Carlo error of empirical design variance \label{appendix:mc_error}}

We use a bootstrap procedure to estimate the Monte Carlo error of the empirical estimate of design variance. For each area $i$ and bootstrap sample $b=1,...,1000$, we define the set of indices, $\mathcal{G}_i^{(b)}$, as a sample from $\mathcal{G}_i$ with replacement, such that $|\mathcal{G}_i^{(b)}|=|\mathcal{G}_i|$. Then we obtain the relative Monte Carlo error, $$\frac{\sqrt{\frac{1}{1000}\sum_{b=1}^{1000}\left(V_i^{(b)}-\frac{1}{1000}\sum_{b'=1}^{1000}V_i^{(b')}\right)^2}}{V_i}$$ where $V_i$ is the empirical design variance defined in (\ref{eq:emp_var}) and $$V_i^{(b)}=\frac{1}{|\mathcal{G}_i|}\sum_{g\in \mathcal{G}_i^{(b)}}\left(\hat\theta_i^{(g)}-\frac{1}{|\mathcal{G}_i|}\sum_{g'\in \mathcal{G}_i^{(b)}}\hat\theta_i^{(g')}\right)^2.$$

\begin{figure}[h]
    \centering
    \includegraphics[width=1\linewidth]{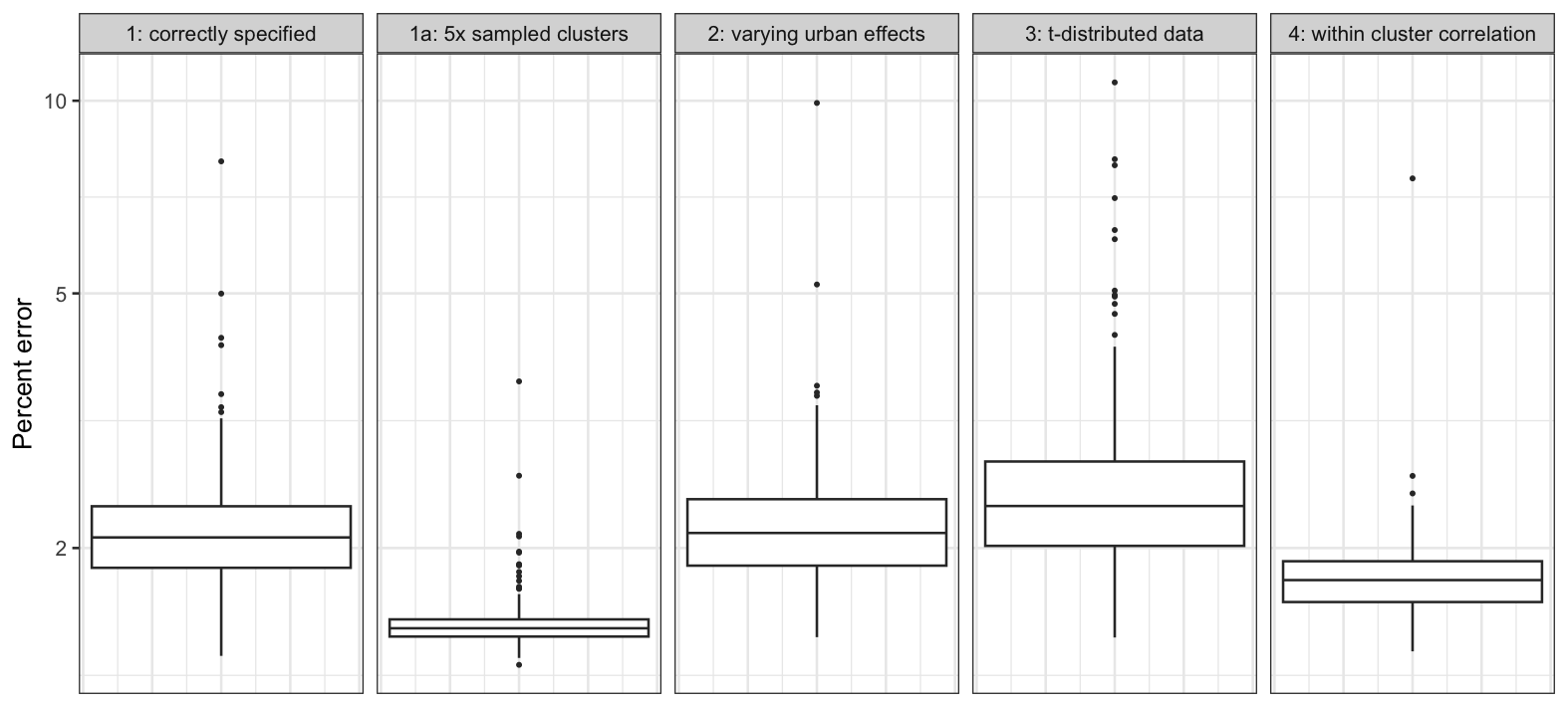}
    \caption{Relative Monte Carlo error of empirical estimate of the true design variance, for each area and setting, estimated using 1000 bootstrap replicates.}
    \label{fig:mc_emp}
\end{figure}

In Figure \ref{fig:mc_emp} we observe that the relative Monte Carlo error is fairly low across all areas and settings. The areas and settings with higher Monte Carlo error are a result of a smaller size $\mathcal{G}_i$. Recall that $|\mathcal{G}_i|\leq 10000$ is determined by the number of sampled datasets (out of $10000$) for which area $i$ has at least $2$ sampled clusters. Thus, areas with fewer clusters will have a higher Monte Carlo error.

\subsection{Comparing properties of sampling distributions and empirical distribution \label{appendix:moments}}

We display the mean, variance, and skewness of the design variance estimator under each sampling distribution for each simulation setting. 

\begin{figure}[h!]
    \centering
    \includegraphics[width=1\linewidth]{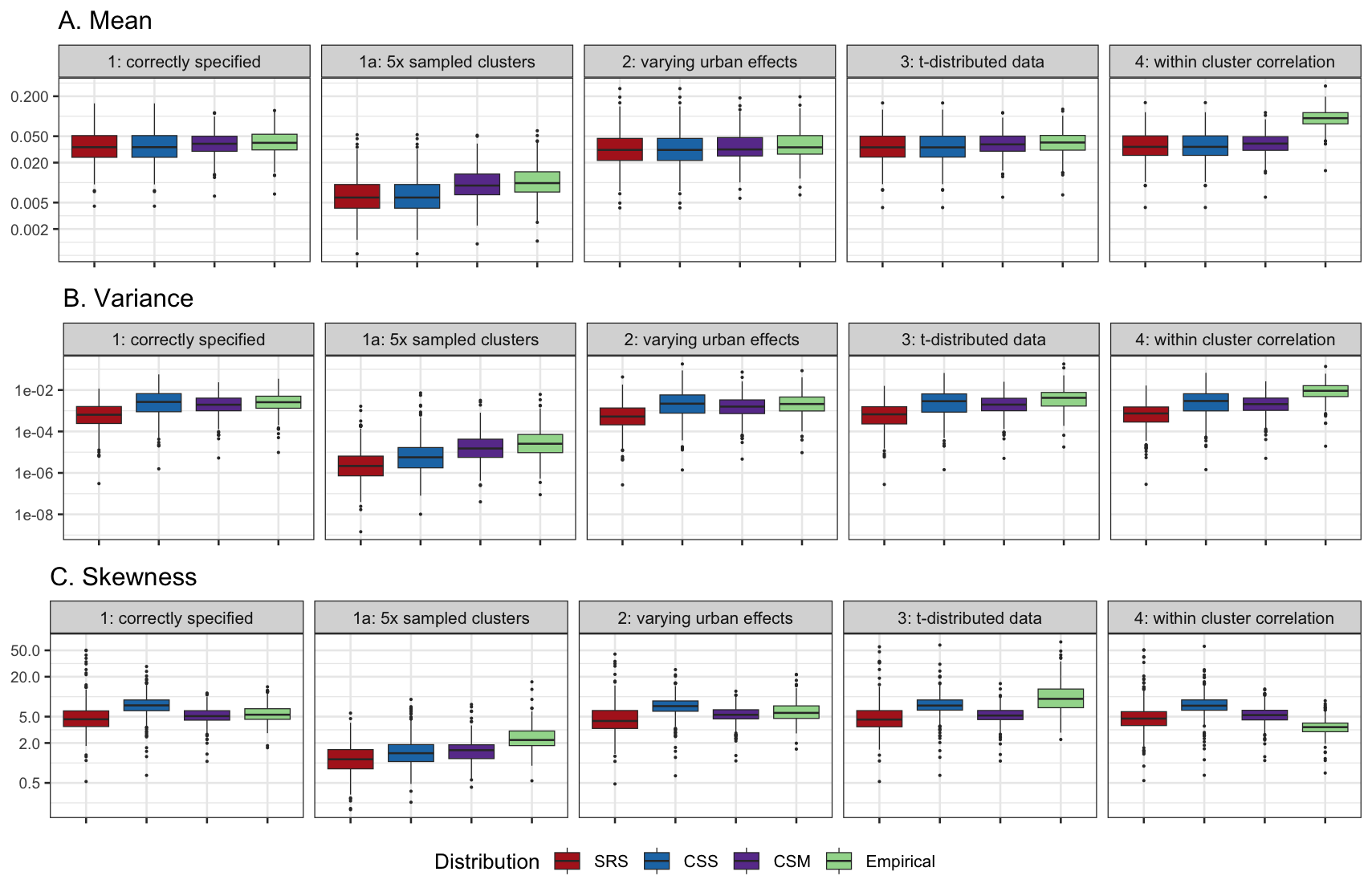}
    \caption{Mean, variance, and skewness of the design variance estimator under each sampling distribution for each area, under 5 simulation settings.}
\end{figure}

\subsection{Re-enumerated cluster sizes \label{appendix:sim_additional}}

\begin{figure}[h!]
    \centering
    \includegraphics[width=1\linewidth]{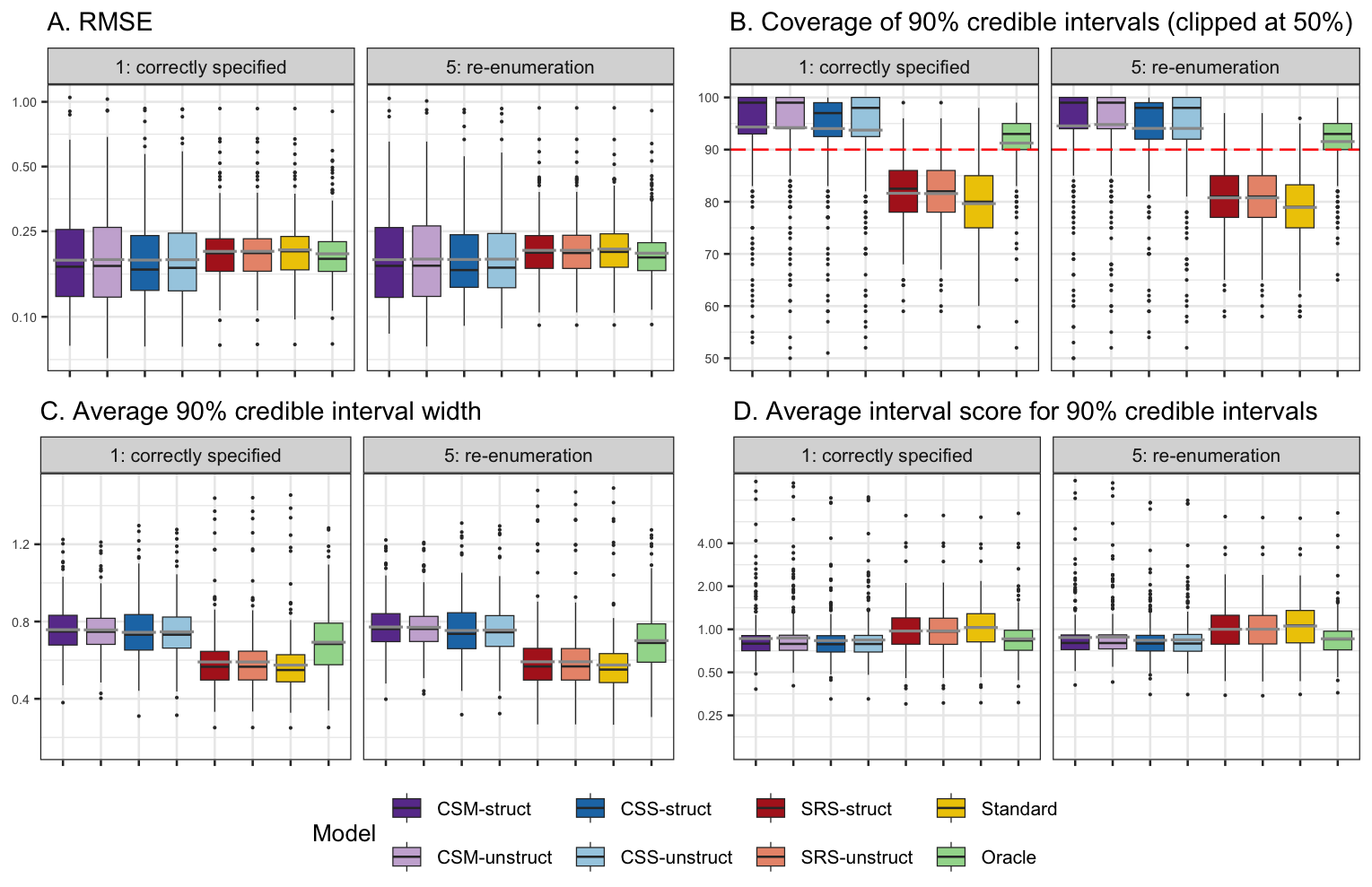}
    \caption{Boxplots of average RMSE of point estimates and coverage, average width, and average interval score of 90\% credible intervals for $300$ areas, for each model and setting. The average for each area is taken across $100$ simulations. The gray lines indicate the mean across areas.}
    \label{fig:sim_additional}
\end{figure}

In the simulation design described in Subsection \ref{sec:sim_design} we assume the listed cluster sizes are correct and do not require re-enumeration (i.e. $L_c=N_c$). In this section we evaluate whether cluster re-enumeration, and the resulting difference in sampling weights, significantly modifies the results we observe in the main text. We use the same setup and parameters here as in setting 1, with the modification that each cluster $c$, is sampled with probability proportional size using $L_c$ (instead of $N_c$) where $L_c$ is generated from a normal distribution with mean $0.85N_c$ and standard deviation $0.2 N_c$, rounded to the nearest whole number. The mean of $L_c$ is chosen to reflect the fact that, due to population growth, we expect the listed cluster size to underestimate the true size, on average. In Figure \ref{fig:sim_additional} we observe that all conclusions made in the main text also hold for this setting.

\clearpage

\setcounter{page}{1}

\section{Distribution of cluster sizes in 2022 Kenya DHS \label{sample_size_appendix}}
We present the distribution of cluster sample sizes for urban and rural clusters in the 2022 Kenya DHS, which we use to calibrate the sample sizes in our simulation study. We use a negative binomial distribution with the parameterization,
\[
P(n_c|\mathrm{size}=s,\mathrm{mean}=\mu)=\left(\begin{matrix}n_c+s-1\\ n_c\end{matrix}\right)\left(\frac{\mu}{\mu + s}\right)^{n_c}\left(\frac{s}{\mu+s}\right)^s.
\]

\begin{figure}[h!]
    \centering
    \includegraphics[width=0.9\linewidth]{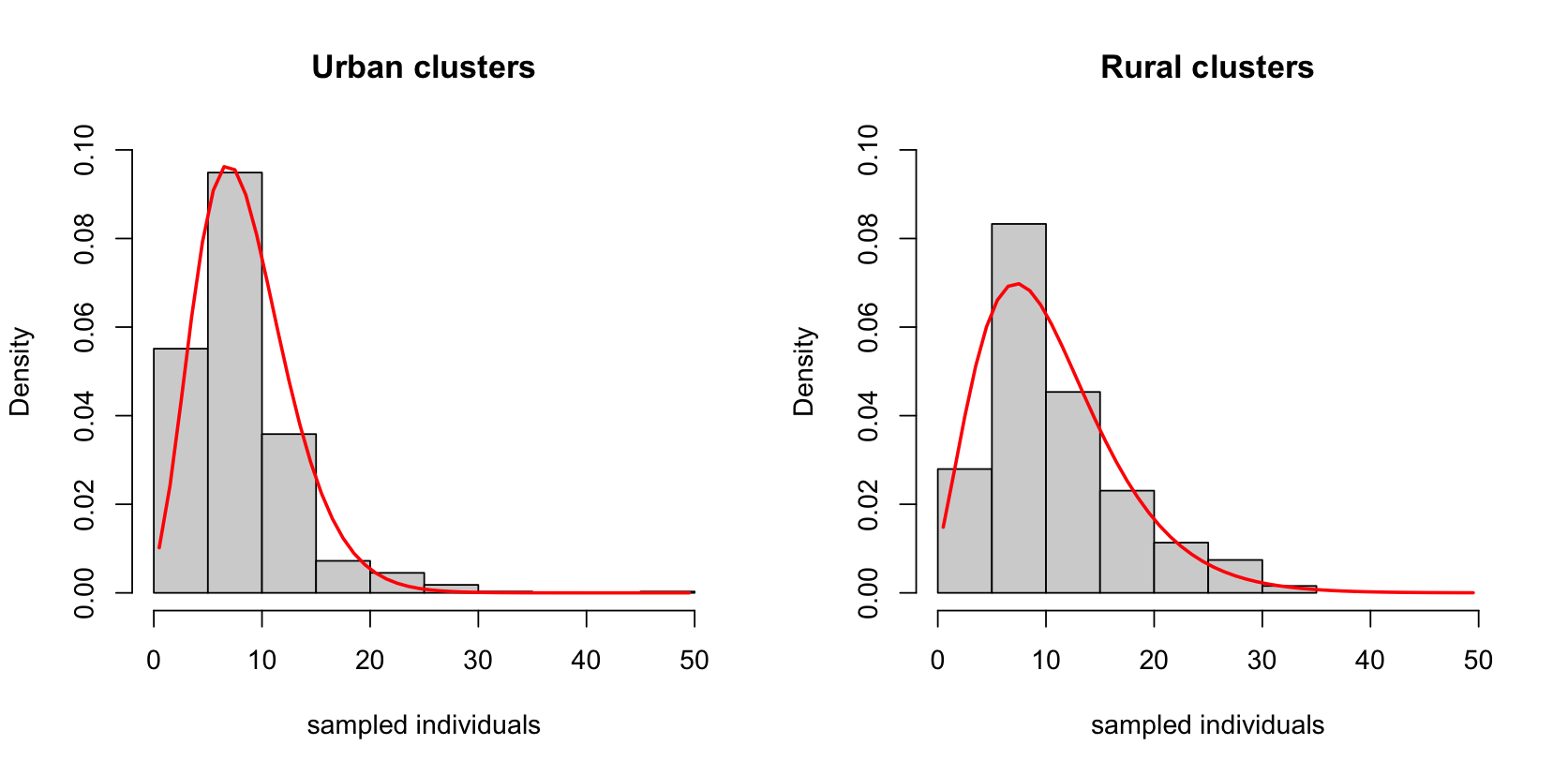}
    \caption{Distribution of urban and rural cluster sample sizes in the 2022 Kenya DHS. The urban cluster size distribution is compared to a negative binomial distribution, denoted by a red curve, with $\mathrm{size}=8$ and $\mathrm{mean}=9$. Similarly, the rural cluster size distribution is compared to a negative binomial distribution, with $\mathrm{size}=4$ and $\mathrm{mean}=11$.}
    \label{fig:cluster_dist_appendix}
\end{figure}

\clearpage

\setcounter{page}{1}
\section{Prior sensitivity \label{appendix:prior_sensitivity}}

\begin{figure}[h]
    \centering
    \includegraphics[width=1\linewidth]{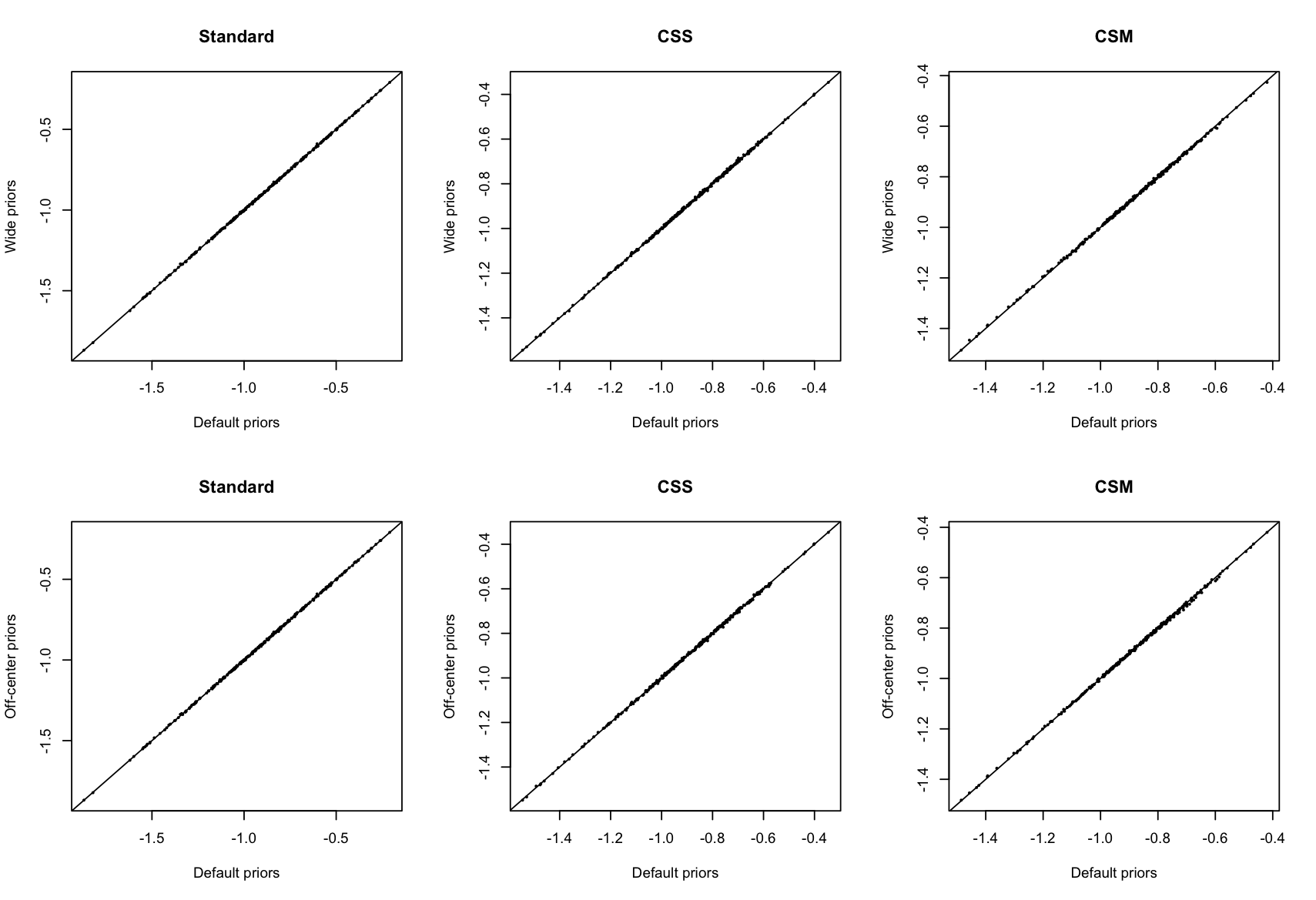}
    \caption{Mean estimates for HAZ using 2022 Kenya DHS under a standard Fay-Herriot model and CSS and CSM variance smoothing models, compared across different prior choices.}
    \label{fig:prior_sens_mean}
\end{figure}

\begin{figure}[h]
    \centering
    \includegraphics[width=1\linewidth]{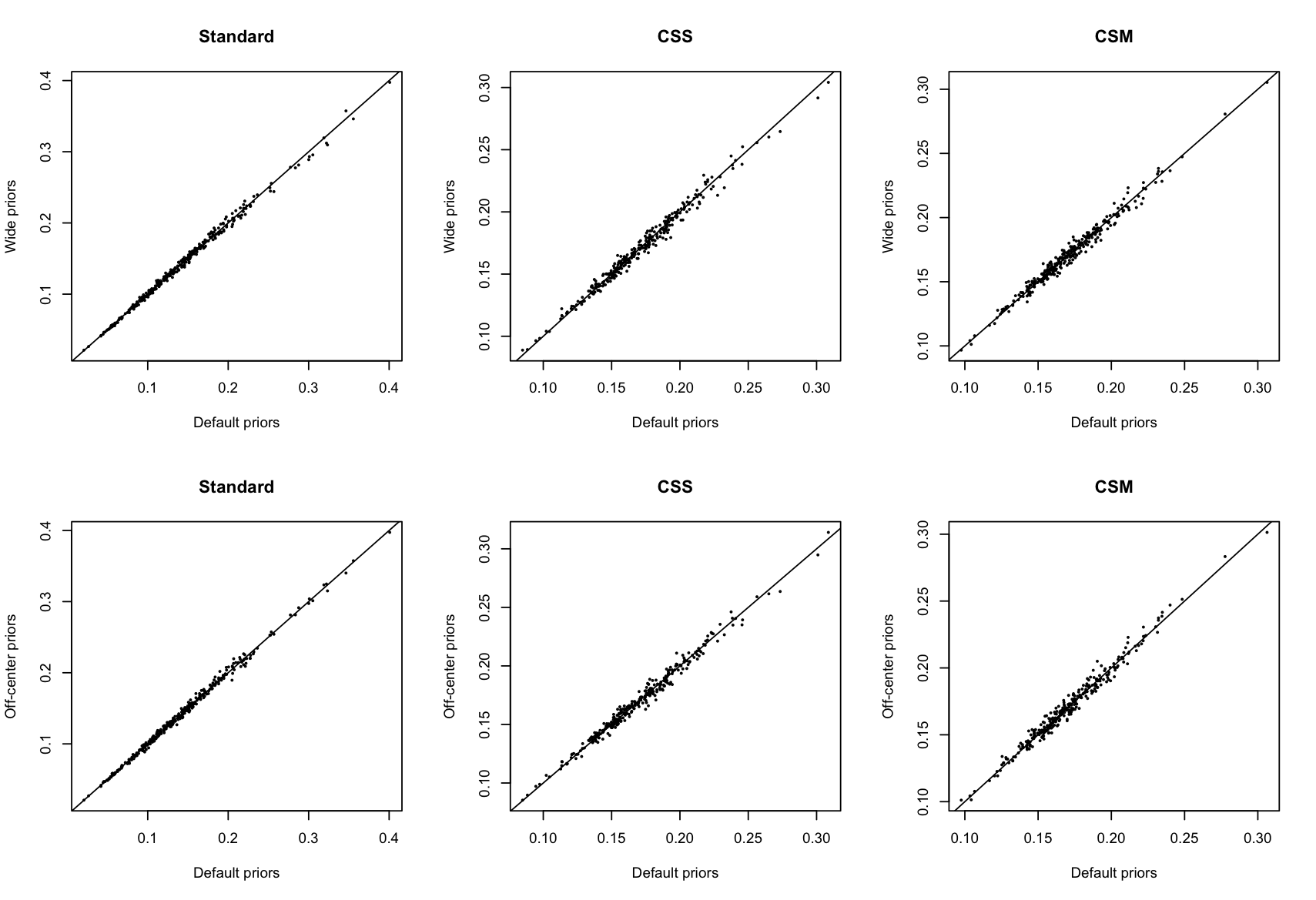}
    \caption{Standard deviation estimates for HAZ using 2022 Kenya DHS under a standard Fay-Herriot model and CSS and CSM variance smoothing models, compared across different prior choices.}
    \label{fig:prior_sens_sd}
\end{figure}

We test the robustness of our prior choices for regression parameters $(\boldsymbol{\beta},\gamma,\boldsymbol{\eta})$ in the standard, CSS, and CSM Fay-Herriot models using the example of HAZ estimation in Kenya. Recall from Sections \ref{sec:candidate} and \ref{sec:data_example} that in both the simulation study and application to the Kenya DHS we choose $\boldsymbol{\beta}\sim\mathcal{N}(0,2^2)$, $\gamma\sim\mathcal{N}(0,2^2)$ and $\boldsymbol{\eta}\sim\mathcal{N}(0,1^2)$. In Figures \ref{fig:prior_sens_mean} and \ref{fig:prior_sens_sd} we compare estimation under these priors to estimation under two different sets of priors. First, we compare to wider priors, $\boldsymbol{\beta}\sim\mathcal{N}(0,5^2)$, $\gamma\sim\mathcal{N}(0,5^2)$, and $\boldsymbol{\eta}\sim\mathcal{N}(0,5^2)$. Second, we compare to priors that have a mean shifted from $0$, $\boldsymbol{\beta}\sim\mathcal{N}(1,2^2)$, $\gamma\sim\mathcal{N}(1,2^2)$, and $\boldsymbol{\eta}\sim\mathcal{N}(0.5,1^2)$. Through these comparisons we observe that the posterior mean and variance estimates for all models are robust to a sensible choice of prior.

\clearpage
\setcounter{page}{1}

\section{Auxiliary variables for Kenya example \label{appendix:kenya_covariates}}

For the area-specific auxiliary variables in the latent mean model we use the urban under-5 population proportion, population density, yearly average temperature, yearly total precipitation, elevation, motorized travel time to healthcare, and yearly average nighttime lights. We only use auxiliary variables which are calculated from satellite data because the inclusion of survey-estimated covariates would introduce non-negligible uncertainty which is outside the scope of this paper.

The urban under-5 population proportions are estimated using the method in \cite{wu}, a logistic classification model which uses pixel-level population density obtained from WorldPop (\citealp{worldpop}) and defines a threshold for classifying pixels as urban or rural, which is calibrated by the urban/rural composition at the Admin-1 level reported by DHS (\citealp{dhs}).  High-resolution surfaces for yearly average temperature, elevation, and yearly total precipitation were obtained from WorldClim (\citealp{worldclim}). High-resolution surfaces for yearly average nighttime lights and motorized travel time to healthcare was obtained from \cite{nighttime_lights} and \cite{time_to_healthcare}, respectively. For each high-resolution surface, we obtained the area-specific auxiliary variable by taking the population-weighted average of the value over each area. We took the log transformation of variables with skewed distributions (population density, nighttime lights, and time to healthcare) and then standardized each variable (except for the urban proportion).

\begin{figure}[h!]
    \centering
    \includegraphics[width=1\linewidth]{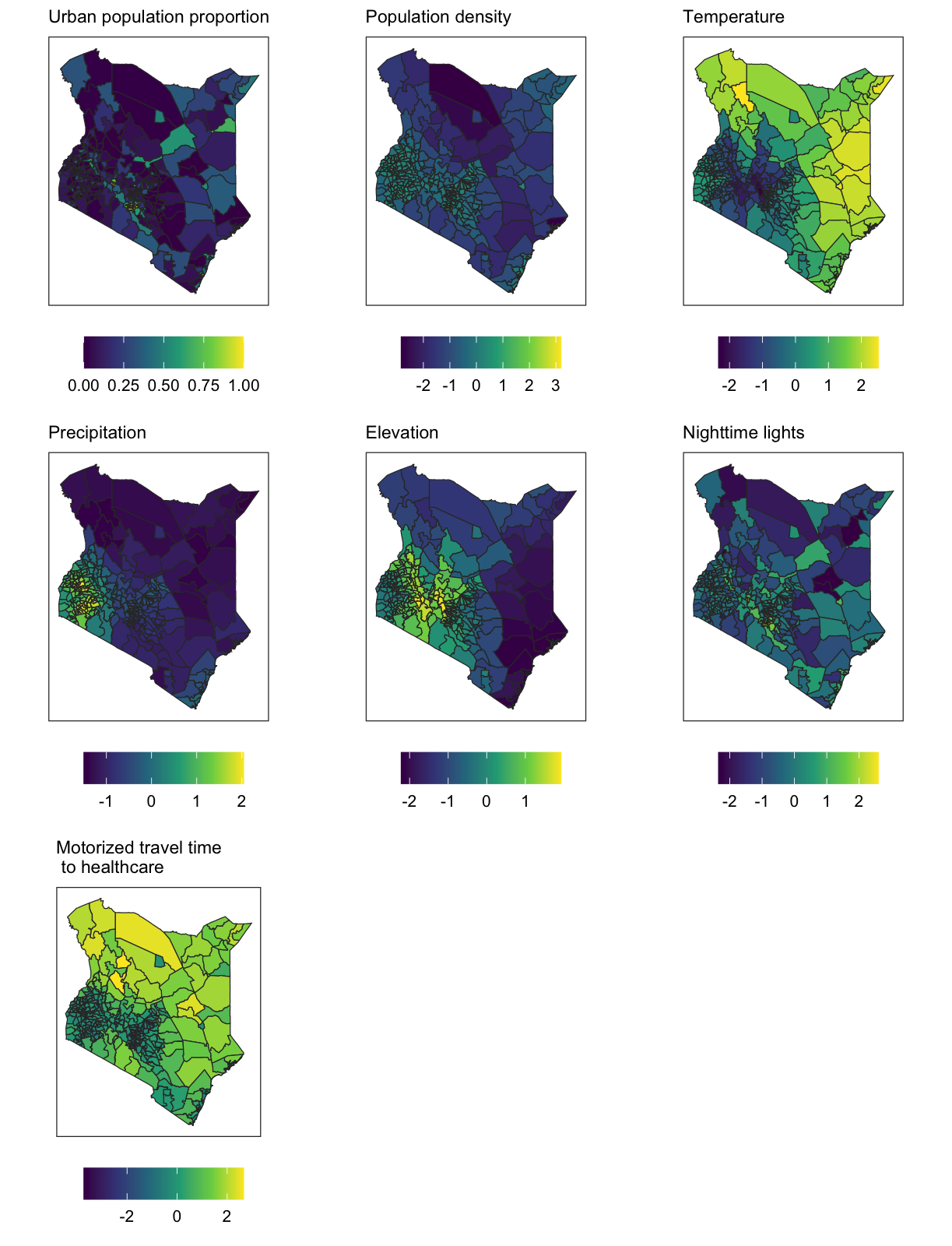}
    \caption{{\bf Maps of standardized Admin-2 area-level auxiliary variables in Kenya}}
\end{figure}

\clearpage

\setcounter{page}{1}
\section{Parameter estimates for Kenya example \label{app:parameter_estimates}}

\begin{table}[h]
    \centering
    \rowcolors{3}{gray!15}{white}
    \begin{tabular}{c|c|c|c|c}
        Parameter & Description & Standard & CSS & CSM \\
        \hline\hline
        \multicolumn{5}{l}{Latent mean model} \\
        \hline\hline
       $\beta_1$ & Intercept  & $-0.95$ & $-0.93$ & $-0.90$ \\
        &   & $(-1.03, -0.88)$ & $(-1.002, -0.85)$ & $(-0.95, -0.84)$\\
       $\beta_2$ & Population density & $-0.04$ & $-0.002$ & $0.02$ \\
        &   & $(-0.13, 0.05)$ & $(-0.10, 0.09)$ & $(-0.08, 0.11)$ \\
       $\beta_3$ & Nighttime lights & $-0.06$ & $-0.05$ & $-0.03$ \\
        &   & $(-0.14, 0.01)$ & $(-0.13, 0.03)$ & $(-0.10, 0.04)$ \\
       $\beta_4$ & Time to healthcare & $-0.03$ & $-0.01$ & $-0.01$\\
        &   & $(-0.12,0.06)$ & $(-0.11, 0.09)$ & $(-0.11, 0.10)$ \\
       $\beta_5$ & Precipitation & $0.02$ & $-0.01$ & $-0.02$ \\
        &   & $(-0.03, 0.07)$ & $(-0.06, 0.04)$ & $(-0.07, 0.03)$ \\
       $\beta_6$ & Temperature & $-0.09$ & $-0.09$ & $-0.07$ \\
        &   & $(-0.23, 0.04)$  & $(-0.02, 0.05)$ & $(-0.22, 0.07)$ \\
       $\beta_7$ & Elevation & $-0.10$ & $-0.11$ & $-0.10$ \\
        &  & $(-0.24, 0.03)$ & $(-0.24, 0.03)$ & $(-0.25, 0.04)$ \\
       $\gamma$ & Urban effect & $0.28$ & $0.16$ & $0.043$ \\
        &   & $(0.01, 0.56)$ & $(-0.12, 0.44)$ & $(-0.14, 0.22)$ \\
       $\sqrt{1/\tau_b}$ & BYM2 standard deviation & $0.36$ & $0.28$ & $0.24$ \\
        &  &  $(0.32, 0.40)$ & $(0.24, 0.32)$ & $(0.20,0.29)$\\
       $\phi_b$ &  BYM2 spatial dependence & $0.75$ & $0.89$ & $0.88$ \\
        & & $(0.54, 0.90)$ & $(0.72, 0.98)$ & $(0.70, 0.97)$ \\
        \hline\hline 
        \rowcolor{white} \multicolumn{5}{l}{Latent variance model} \\
       \hline\hline
       \rowcolor{gray!15} $\eta$ & Intercept &  - & $1.19$ & $1.44$ \\
       \rowcolor{white} &   &  & $(1.10, 1.28)$ & $(1.34, 1.54)$ \\
       \rowcolor{gray!15} $\sqrt{1/\tau_g}$ & BYM2 standard deviation & - & $0.13$ & $0.08$ \\
       \rowcolor{white} &   &  & $(0.01, 0.31)$ & $(0.01, 0.25)$ \\
       \rowcolor{gray!15} $\phi_g$ & BYM2 spatial dependence & - & $0.26$ & $0.32$ \\
       \rowcolor{white} & & - & $(0.02, 0.75)$ & $(0.03, 0.79)$ \\
     \end{tabular}
     \caption{Parameter estimates under each model estimating HAZ using 2022 Kenya DHS data. 90\% credible intervals are in parantheses.}
\end{table}

\clearpage
\setcounter{page}{1}

\section{Scatter plots of Kenya HAZ estimates \label{app:kenya_scatter}}

In this section we compare the mean and standard deviation estimates of HAZ under each model to the design-based estimates. As in Section \ref{sec:data_example} we observe that both variance smoothing models exhibit more shrinkage of the mean and variance compared to the standard model. It is important to note that this shrinkage is not necessarily a negative characteristic, as the design-based estimates are quite noisy due to low sample size. The design-based standard deviation estimates, in particular, are not very reliable, as is noted throughout the main text and exhibited in Subsections \ref{section:var_est_performance} and \ref{section:mean_est_performance} of the simulation study.

\begin{figure}[h!]
    \centering
    \includegraphics[width=1\linewidth]{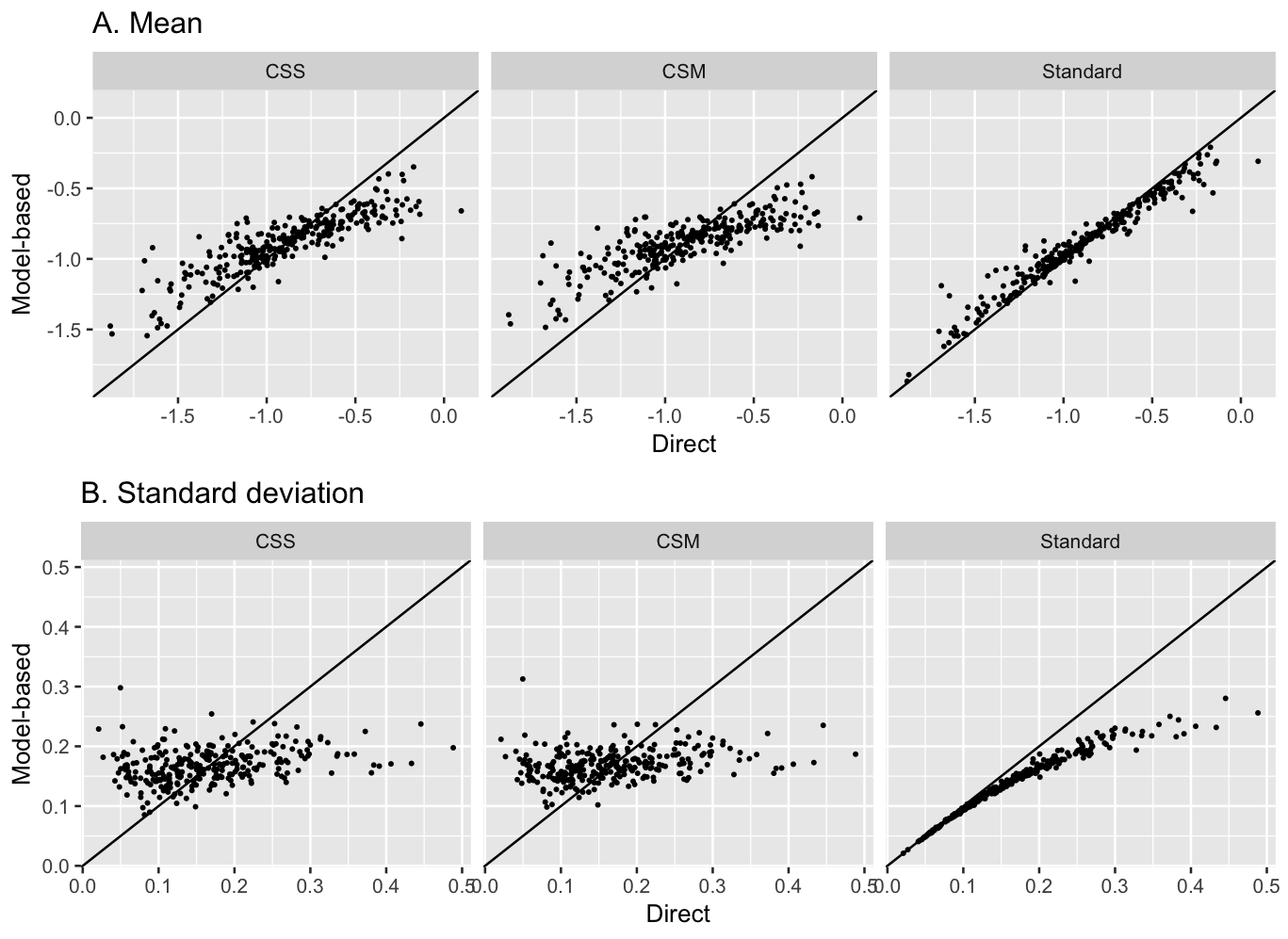}
    \caption{Scatter plots comparing mean and standard deviation estimates of HAZ under each model against the design-based estimates.}
\end{figure}

\clearpage

\end{document}